\documentclass[12pt,amssymb]{iopart}  \usepackage{graphicx}
\usepackage{subfigure,cite}

\begin{document}

\title[Response of a Hexagonal Granular Packing]{Response of a
Hexagonal Granular Packing under a Localized  External Force: Exact
Results}

\author{Srdjan Ostojic and Debabrata Panja}

\address{Institute for Theoretical Physics, Universiteit van
Amsterdam, Valckenierstraat 65, 1018 XE Amsterdam, The Netherlands}

\begin{abstract} 

We study the response of a two-dimensional hexagonal packing of
massless, rigid, frictionless spherical grains due to a vertically
downward point force on a single grain at the top layer. We use a
statistical approach, where each mechanically stable configuration of
contact forces is equally likely. We
show that this problem is equivalent to a correlated $q$-model. We find that the response is
double-peaked, where the two peaks, sharp and single-grain diameter
wide, lie on the two downward lattice directions emanating from the
point of the application of the external force. For systems of finite size, the magnitude of these
peaks decreases towards the bottom of the packing, while progressively
a broader, central maximum appears between the peaks. The response
behaviour displays a remarkable scaling behaviour with system size
$N$: while the response in the bulk of the packing scales as
$\frac{1}{N}$, on the boundary it is independent of $N$, so that in
the thermodynamic limit only the peaks on the lattice directions
persist. This qualitative
behaviour is extremely robust, as demonstrated by our simulation
results with different boundary conditions. We have obtained exact
expressions of the response and higher correlations for any system
size in terms of integers corresponding to an underlying discrete
structure.

\end{abstract}

\maketitle

\section{Introduction}
Static properties of granular packings have been a very active field
of research in recent years. Granular packings are
assemblies of macroscopic particles that interact only via mechanical
repulsion mediated through physical contacts \cite{review,Brev}. In
contrast with continuum solids, forces on individual grains in a
granular packing can be directly accessed both experimentally
\cite{fluct_exp,exp1,exp2} and numerically
\cite{fluct_num,resp_num,tamas,macnam}. These forces are found to be
organized in highly heterogeneous networks that depend strongly on
construction history of the packing \cite{vanel}. Statistical studies,
motivated to deal with such history dependence and heterogeneity of
the forces on individual grains, have identified two main global
characteristics of static granular packings. First,
the distribution for the magnitudes of inter-grain forces is very
broad, with an exponential decay for large force magnitudes and a
plateau at small force magnitudes  \cite{fluct_exp,
fluct_num}. Secondly, the average response of the packings to a single
vertically downwards external force depends strongly on the underlying
geometry: in ordered packings, the applied force is mainly transmitted
along the principal lattice directions emanating from the point of
application of the force, while in disordered packings there is a
single vertical propagation direction \cite{exp1,exp2,resp_num}.

Although a large number of different theoretical models have been
proposed to study these two global characteristics of granular
packings, each of these models successfully accounts for at most one of
them. For example, the so-called $q$-model \cite{qmodel} is a scalar
lattice model that describes the fluctuations in force transmission
within a granular packing at the first approximation. While this model
does produce realistic distributions for the magnitudes of inter-grain
forces (see Ref. \cite{jacco1} for a more detailed discussion),  it
predicts a diffusive response to a single external force  in conflict
with experiments \cite{qresponse}. On the other hand, vectorial
extensions of the $q$-model \cite{qresponse}, compatible with a more
general ``stress-only'' approach \cite{stress_only},  were shown to lead
to stochastic wave equations.  For weak disorder, these equations
predict a ray-like propagation of stresses in agreement with
experiments, but for strong disorder, the corresponding behaviour of
the stresses is less clear-cut. This has further led to the introduction
of an ad-hoc ``force chain splitting'' model \cite{chain_split}. The
vectorial extensions of the $q$-model and the force-chain splitting
models do predict realistic response behaviour for granular packings,
but the relation between microscopic stochasticity and resulting
distributions for the magnitudes of inter-grain forces within these
models is not entirely clear. Finally, in contrast with these
stochastic approaches, classical elastic theory has been used for
years in the engineering community \cite{ned, gold}. The predictions of
this theory for the response behaviour of granular packings match
the experimental results rather successfully \cite{elastic}. However,
 elasticity theory provides only a
macroscopic, average level description, and thus it provides no
information on the force distribution. Moreover, its derivation from
the grain-level mechanics still seems to be a significant challenge \cite{ned,gold}.

From this perspective, a unifying approach leading to both realistic
fluctuations in the individual inter-grain force magnitudes and
transmissions of forces in a granular packing clearly seems to be
necessary. Interestingly, in a different context (namely, that of
granular compaction) a basis for such an approach has been laid down
by Edwards years ago \cite{edwards, edwards2}. By analogy with
conventional statistical mechanics, the fundamental hypothesis
is to consider all ``jammed'' configurations equally likely. Although
no microscopic justification for such an ergodic assumption is
available so far,  experiments and numerical studies of
quasi-static granular media support such a thermodynamic picture
\cite{jamming}. It thus seems very tempting to extend this hypothesis
to the study of forces in static granular packings by considering sets
of forces belonging to all mechanically stable configurations equally
likely.

If this idea of equal probability ensemble is to be applied to study
the forces in a granular packing, it has to be realized that two
levels of randomness are generally present in the ensemble of forces
for stable granular packings \cite{Brev}. This stems from the fact that forces in
a granular packing depend critically on the underlying contact network
between the grains. Contact networks that are deemed isostatic
uniquely determine the forces admissible on them, and thus, the application of the equal-probability hypothesis to packings with
isostatic contact networks amounts to considering each contact network
equally likely. Such a consideration leads to wave equations in
disordered geometry whach are in conflict with experiments \cite{edw3}. However,
realistic contact networks are generically non-isostatic, and the fact
that several force configurations can be compatible with a given
non-isostatic contact network gives rise to the second level of
randomness \cite{tamas,macnam,silb}.

\begin{figure}[h]
\begin{center} 
\includegraphics[width=0.9\linewidth]{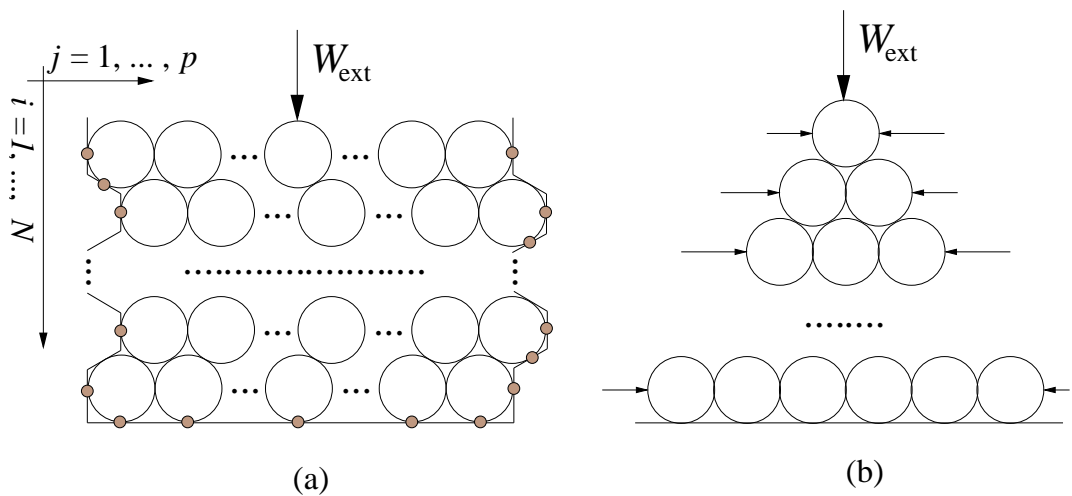}
\caption{Our model: (a) $N\times p$ array of hexagonally close-packed
rigid frictionless spherical grains in two-dimensions.  At the top,
there is only a single vertically downward point force applied on one
grain. (b) For massless grains, the
vertical forces are non-zero only inside a triangle formed by the two
lattice directions emanating from the point of application of
$W_{\mathrm{ext}}$, so that we can restrict our study to that
domain. The horizontal forces on boundaries of the triangular domain are the same
as those on the boundaries in (a).}
\label{fig1}
\end{center}
\end{figure}

Instead of applying Edward's hypothesis to both levels of randomness
simultaneously, a natural first step is to apply the uniform
probability hypothesis first to a fixed contact network, and then
possibly average over various contact networks. Such a
study in a fixed contact network has recently been shown to produce
distributions for the magnitudes of inter-grain forces that compare
very well with experiments \cite{jacco2}. In Ref. \cite{art1} we
briefly showed that the application of the uniform probability
hypothesis in an ordered geometry also leads to a response to an
external force qualitatively in agreement with experiments. In this
paper, we report the same study in full detail. More precisely, we
determine the behaviour of the response of a two-dimensional hexagonal
packing of rigid, frictionless and massless spherical grains placed
between two vertical walls (see Fig.\ref{fig1}), due to a vertically
downward force $W_{\mathrm{ext}}$ applied on a single grain at the top
layer. We define the response of the packing as $\left[\langle
W_{i,j}\rangle -\langle W_{i,j}^{(0)}\rangle\right]/W_{\mathrm{ext}}$,
where $W_{i,j}$ and $W_{i,j}^{(0)}$ are the vertical forces
transmitted by the $(i,j)$-th grain to the layer below it respectively
with and without applied external force $W_{\mathrm{ext}}$. For
massless grains $W^{(0)}_{i,j}\equiv0$. The angular brackets here
denote averaging with equal probability over all configurations of
mechanically stable non-negative contact forces.

We find that the problem of equally-probable force configurations in
the hexagonal geometry is equivalent to a correlated $q$-model. Using
this formulation, simulations for a variety of boundary conditions
show that the response to an external force applied on the top of the
packing displays two symmetric peaks lying precisely on the two
downward lattice directions emanating from the point of application of
the force. Moreover, the response exhibits remarkable scaling with the
system size, implying that in the thermodynamic limit the peaks
propagate all the way to the bottom of the packing. Surprisingly,
average values $\langle W_{i,j}\rangle$ as well as higher correlations
can be calculated {\it exactly\/} for any system size via this
formulation. We show that all these quantities can be expressed in
terms of integers corresponding to an underlying discrete structure.

The paper is organized as follows. In Sec. \ref{sec2}, we examine the
equal-probability hypothesis in a fixed geometry, and its application
to the hexagonal geometry in detail. In Sec. \ref{sec3}, we numerically
study the response as well as the distributions of the single $q$'s
and correlations between them. In Sec. \ref{sec4} we detail the full
theoretical calculation of $\langle W_{i,j}\rangle$ and higher
correlations for any system size. We finally end this paper with a
discussion in Sec. \ref{sec5}.

\section{Uniform Measure on the Force Ensemble \label{sec2}}

\subsection{Force Ensemble in Generic Packings\label{sec2.1}}

To start with, let us examine more closely the inter-grain forces in a
granular packing \cite{roux}.
As pointed out earlier, the grains
mutually interact only via (mechanical) repulsive contact forces. The
contacts are assumed pointlike, and in the present  study, we consider
only frictionless grains, so that each contact force is locally normal
to the grain surface.  Thus, the most fundamental entity entering the
description of forces in a granular packing is the {\it contact
network}, i.e., the set of all contact points with directions normal
to the grain surfaces at respective contacts \cite{edwards2}.

Consider such a contact network formed by a packing of $P$ grains with
$Q$ contacts. For simplicity, we restrict ourselves to two dimensions
in the following analysis. At each contact $k$ for $1 \leq k \leq Q$,
the force is represented by its magnitude $F_{k}$ along the locally
normal direction to the grain surface, with the convention that a
positive magnitude of $F_{k}$ corresponds to a repulsive force.  The
forces applied on a given grain must satisfy three Newton's equations:
two for balancing forces in the $x$ and $y$ directions and one for
balancing torque. For the system in its entirety, the contact forces
can be grouped in a column vector $\mathsf F$ consisting of $Q$
non-negative scalars $\{F_{k} \}$, satisfying $S=3P$ stability
equations represented by ${\mathbf  A}\cdot{\mathsf F}={\mathsf
F}_{\mathrm{ext}}$. Here, $\mathsf{F}_{\mathrm{ext}}$ is an
$S$-dimensional column vector representing the external forces and
torques on the grains, and $\mathbf{A}$ is an $S\times Q$ matrix
uniquely specified by the contact network. If the grains in question
are disks, as is the case in most of theoretical and numerical
studies, the torque balance is automatically  satisfied, and the total
number of equations $S$ drops to $2P$. Note also that there is some
freedom in the definition of $\mathsf F$: a contact force between a
grain and a boundary can either be considered as an internal force,
i.e., as a part of $\mathsf F$; or as an external force, in which case
it becomes a part of $\mathsf{F}_{\mathrm{ext}}$.

In the above description, if ${\mathbf A}$ is invertible --- which of
course implies $Q=S$ --- the force configuration allowed on the
contact network is unique: such a contact network is called
isostatic. Otherwise either there is an extended set ${\cal E}$ of
allowed force configurations, in which case the network is called
hyperstatic, or there is none, the packing under consideration is
unstable under the imposed external forces.

Theoretical arguments suggest that perfectly rigid grains generically
form isostatic packings \cite{mouk}.  Physical grains are of course
never infinitely rigid, and the corresponding contact networks are
hyperstatic. These observations are confirmed by numerical studies,
which moreover find that the convergence to isostaticity with
increasing rigidity of the grains is rather slow \cite{silb}.  To
determine the forces uniquely, one then in principle has to take into
account the ``force law'', which relates the deformation of a particle
to the force applied on it, as well as the construction history of
the packing.

Nevertheless,
some macroscopic properties of a granular packing, such as the
ditribution of force magnitudes and the shape of the average response
 to a point force, are independent of
the details of the force law \cite{jacco2}. From that point of view, if 
one
considers a packing of grains of large but finite rigidity, the
deformations of the grains are small with respect to their
characteristic sizes, and these deformations could be altered without 
essentially
modifying the (hyperstatic) contact network \cite{jacco2}. Experimentally, 
this can
be achieved by gently tapping the packing without adding or removing
contacts \cite{Brev}.  Without having to delve deeper in the
microscopic force laws, one can then analyze the contact force
configurations  as a subset of the set ${\cal E}$ of allowed force
configurations. By analogy to classical statistical mechanics, one can
 study the statistical properties of the set ${\cal E}$. A natural
starting point is to assume that any point in ${\cal E}$ is visited
with the same frequency, similar to a microcanonical ensemble.  In
other words, one assigns a uniform probability measure on ${\cal E}$,
under which all allowed contact force configurations are equally
likely. We should however keep in mind that {\it a priori\/} there is no
clear justification for such an ergodic hypothesis.

The uniform measure on ${\cal E}$ can be defined more precisely.  The
set of all solutions of ${\mathbf A}\cdot{\mathsf F}={\mathsf
F}_{\mathrm{ext}}$ is an affine space, whose dimension is the
dimension  of the kernel of ${\mathbf A}$, and ${\cal E}$ is a subset
of that space where $F_{k} \geq 0 \,\, \forall k=1 \ldots Q$. It can
be shown that ${\cal E}$ is usually a compact polygon \cite{tamas}, so
that the uniform measure is well defined.  For a hyperstatic packing,
$Ker(A)$ is  non-zero, and a parametrization of ${\cal E}$  can be
constructed  via the three following steps: (1)  one first identifies
an orthonormal basis $\{{\mathsf F}^{(l)}\}$ ($l=1,\ldots,d_K=Q-S$)
that spans the space of $Ker({\mathbf A})$; (2) one then determines a
unique solution ${\mathsf F}^{(0)}$ of ${\mathbf A}\cdot{\mathsf
F}^{(0)}={\mathsf F}_{\mathrm{ext}}$  by requiring ${\mathsf
F}^{(0)}.{\mathsf F}^{(l)}=0$ for $l=1,\ldots,d_K$; and (3) one
finally obtains all solutions of ${\mathbf A}\cdot{\mathsf F}={\mathsf
F}_{\mathrm{ext}}$ as ${\mathsf F}={\mathsf
F}^{(0)}+\sum\limits_{l=1}^{Q-S}f_l\,{\mathsf F}^{(l)}$, where $f_l$
are real numbers. The restriction of the $f_l$'s  to a set ${\cal S}$
generating non-negative forces is a possible parametrization of ${\cal
E}$.  The uniform measure on ${\cal E}$ is thus equivalent to the
uniform measure $d\mu=\prod\limits_{k} dF_k\, \delta({\mathbf
A}\cdot{\mathsf F}-{\mathsf F_{\mathrm{ext}}}
)\,\Theta({F_k})=\prod\limits_{l} df_l$ on ${\cal S}$.

\subsection{Force Ensemble in the Hexagonal Geometry \label{sec2.2}}

\begin{figure}[h]
\begin{center}
\includegraphics[width=0.9\linewidth]{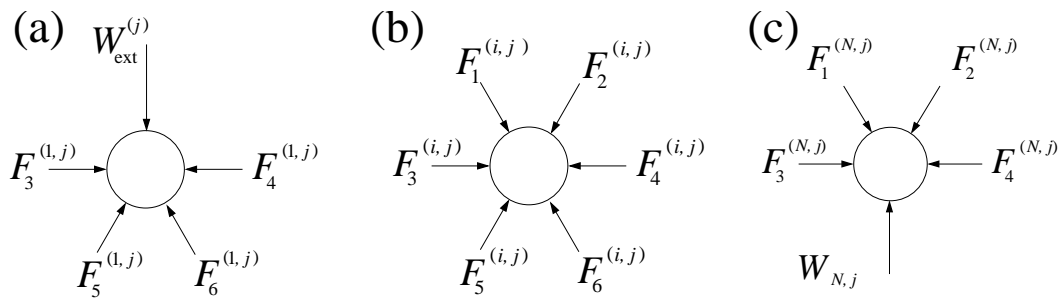}
\caption{Schematically shown forces on the $j$-th grain in the $i$-th
layer: (a) $i=1$, (b) $i\leq N$ and (c) $i=N$;
$F^{(i,j)}_m\geq0\,\,\forall m$.}
\label{fig2}
\end{center}
\end{figure}

We will now apply the general method described in Sec. \ref{sec2.1} to
a hexagonal packing of monodisperse, rigid and  frictionless disks
under the localized external force $W_{\mathrm{ext}}$
(see. Fig. \ref{fig1}).  We will concentrate on massless particles
since considering masses does not fundamentally change the qualitative
behaviour \cite{art1}.  Our aim is to calculate the mean of
$W_{i,j}=\frac{\sqrt{3}}{2}\left[F^{(i,j)}_1+F^{(i,j)}_2\right]$ (see
Fig. \ref{fig2}) by considering all admissible force configurations to
be equally likely.

\subsubsection{Force balance on individual grains:}

To start with, we calculate the forces $F^{(i,j)}_5$ and $F^{(i,j)}_6$
for each grain in terms of other forces by using Newton's
equations. For the top layer, i.e., $i=1$ [Fig. \ref{fig2}(a)],
\begin{eqnarray}
F^{(1,j)}_5=\frac{1}{\sqrt{3}}W_{\mathrm{ext}}^{j}+[F^{(1,j)}_4-F^{(1,j)}_3]\nonumber\\
F^{(1,j)}_6=\frac{1}{\sqrt{3}}W_{\mathrm{ext}}^{j}-[F^{(1,j)}_4-F^{(1,j)}_3]\,,
\label{force_balance_top}
\end{eqnarray}
while for a grain in the bulk [Fig. \ref{fig2}(b)]
\begin{eqnarray}
F^{(i,j)}_5=F^{(i,j)}_2+[F^{(i,j)}_4-F^{(i,j)}_3]\nonumber\\
F^{(i,j)}_6=F^{(i,j)}_1-[F^{(i,j)}_4-F^{(i,j)}_3]\,,
\label{force_balance_bulk}
\end{eqnarray}
and finally for the bottom layer [Fig. \ref{fig2}(c)],
\begin{eqnarray}
W_{N,j}=\frac{\sqrt{3}}{2}\left[F^{(N,j)}_1+F^{(N,j)}_2\right]\nonumber\\
F^{(N,j)}_4-F^{(N,j)}_3=\frac{1}{2}\left[F^{(N,j)}_1-F^{(N,j)}_2\right]\,.
\label{force_balance_bottom}
\end{eqnarray}

In our study, a vertically downward force $W_{\mathrm{ext}}$ is
applied on a single grain  $j_0$ in the top layer, i.e.,
$W_{(1,j)}=W_{\mathrm{ext}}\delta_{j, j_0}$. Equations
(\ref{force_balance_top}-\ref{force_balance_bottom}) then show that
this force can propagate only inside a triangle formed by the two
downward lattice directions emanating from the grain $j_{0}$. Outside the 
triangle, all non-horizontal forces
are zero, and for the horizontal ones,
$F^{(i,j)}_3=F^{(i,j)}_4$. Since our main interest are the
$W_{(1,j)}$'s, this implies that we can restrict ourselves to the
triangular domain, for which the forces exerted on the boundaries are
the same as the forces on the boundary of the original system [see
Fig. \ref{fig1}(b)].

\subsubsection{Parametrization of the force ensemble:\label{sec2.2.1}}

As stated in Sec. \ref{sec2.1}, ${\mathsf F}$ contains one scalar
(force magnitude) for each inter-grain contact. This is a consequence
of the action-reaction principle, which gives the following
identifications for $1<i \leq N$ and $1<j \leq i$:
\begin{eqnarray}
F_{1}^{(i,j)}=F_{6}^{(i-1,j-1)},\qquad F_{2}^{(i,j)}=F_{5}^{(i-1,j)},
\qquad \mbox{and}\qquad F_{3}^{(i,j)}=F_{4}^{(i,j-1)}\,.
\end{eqnarray}

We consider two different vertical boundaries: (i)  ``hard walls''
described in Fig. \ref{fig1} and (ii) periodic boundaries in the $j$
direction. While in both cases the contact network is clearly
hyperstatic, the configurations of the internal forces are slightly
different. We start with the case of ``hard walls''.

For reasons which will become clear later on, we will consider the
contact forces on the right and the bottom boundaries as internal
forces, i.e., as a part of ${\mathsf F}$, while the forces on the top
and right boundaries will be considered as a part of
$\mathsf{F}_{\mathrm{ext}}$. Simple counting then shows that
$Q=\frac{3}{2}N^2+\frac{1}{2} N$,
while the number of equations is $N(N+1)$, implying that
$d_K=\frac{N(N-1)}{2}$.

Notice from Eqs. (\ref{force_balance_top}-\ref{force_balance_bulk})
that in this description, the lateral forces $F^{(i,j)}_3$ and
$F^{(i,j)}_4$ enter the equations for the non-horizontal forces only
via their difference $G_{i,j}=F^{(i,j)}_4-F^{(i,j)}_3$ for $i=1\ldots
N-1$ and $j=1\ldots p$. A natural parametrization of ${\cal E}$ is
given by these $G_{i,j}$'s: once the $G_{i,j}$ are fixed, all the
other non-horizontal forces can be uniquely determined by solving
Eqs. (\ref{force_balance_top}-\ref{force_balance_bottom}) layer by
layer from top down. It is easily seen that the number of these
parameters is indeed $d_K$, and that they correspond to an orthonormal
basis of $Ker{\mathbf A}$.

Thus, the set ${\cal E}$ is parametrized by the $G_{i,j}$'s, and the
$G_{i,j}$'s are restricted within a subset ${\cal S}$ in order to keep all
internal forces non-negative. The
non-negativity of $F_5^{(i,j)}$ and $F_6^{(i,j)}$ implies
\begin{eqnarray}
-\frac{1}{\sqrt{3}}W_{\mathrm{ext}}\leq G_{1,j_0}\leq
\frac{1}{\sqrt{3}}W_{\mathrm{ext}}\qquad \mbox{and}\qquad
-F^{(i,j)}_2\leq G_{i,j} \leq F^{(i,j)}_1\,.
\label{Gineq}
\end{eqnarray}

Furthermore, the horizontal forces can be expressed as
\begin{eqnarray}
F^{(i,j)}_4=F^{(i,1)}_3+{\textstyle\sum\limits_{j'=1}^j}G_{i,j'}\,,
\label{FGtransform}
\end{eqnarray}
where $F^{(i,1)}_3$'s are {\it external} forces. If the magnitudes of
$F^{(i,1)}_3$'s are taken to be larger then
$\frac{2}{\sqrt{3}}W_{\mathrm{ext}}$, then it is easy to see that
$F^{(i,j)}_4$'s remain positive for {\it all\/} values of $G_{i,j}$'s
that satisfy inequality (\ref{Gineq}); otherwise, to have
$F^{(i,j)}_3\geq0$, the available range for each $G_{i,j}$ would
have to be further restricted within the bounds defined by the
inequality (\ref{Gineq}).  In this paper, we  concentrate on the case 
$F^{(i,1)}_3=\frac{2}{\sqrt{3}}W_{\mathrm{ext}}$ where the
full range  (\ref{Gineq}) is allowed. We will consider the case
$F^{(i,1)}_3<\frac{2}{\sqrt3}W_{\mathrm{ext}}$ briefly in
Sec. \ref{sec3.1}.

Notice that the positivity conditions for $F^{(i,j)}_4$'s defined via
Eq. (\ref{FGtransform}) is the only place where the precise values of
$F^{(i,1)}_3$'s enter the analysis. Clearly, considering
$F^{(i,1)}_3$'s as a part of $\mathsf F$ would lead to them being
parameters of ${\cal E}$ (and thus an unbounded ${\cal S}$ due to the
lack of upper bounds of $F^{(i,1)}_3$'s). We therefore choose
$F^{(i,1)}_3$'s as external forces with {\it fixed\/} values
$\frac{2}{\sqrt{3}}W_{\mathrm{ext}}$ to keep ${\cal S}$ bounded and
the uniform measure well-defined.

With the above convention, the set ${\cal S}$ is an
$\frac{(N-1)N}{2}$-dimensional polyhedron in the $G_{i,j}$-space
defined by inequalities (\ref{Gineq}).  The response of the packing to
the external force $W_{\mathrm{ext}}$ is then defined as
\begin{eqnarray}
\langle W_{i,j} \rangle \equiv \frac{1}{{\cal N}}\displaystyle{\int}_{\!\!\!\!\cal S}
\,W_{i,j}\,\prod_{k,l}dG_{k,l};
\label{waverage}
\end{eqnarray}
i.e., the averages $\langle\ldots\rangle$ are defined over an ensemble
of force configurations, where each force configuration is equally
likely, with ${\cal N}=\displaystyle{\int_{{\cal S}}{\prod\limits_{k,l}}\,dG_{k,l}}$ 
the normalization constant.

Most of the previous analysis  remains valid for periodic
boundaries as well. The main difference  is that the $F^{(i,1)}_3$'s
are now internal forces (i.e., a part of $\mathsf F$), as
$F^{(i,1)}_3=F^{(i,p)}_4$ due to the action-reaction principle.  The
phase space ${\cal E}$ can again be parametrized by $G_{i,j}$ for
$i\leq N,1\leq j\leq N$, with the additional constraint that
\begin{eqnarray}
\sum_{j=1}^p G_{i,j}=0 \qquad \forall i=1\ldots N
\label{period}.
\end{eqnarray}
Once again, only the $G_{i,j}$'s physically enter the problem, and the
precise values of the horizontal forces are immaterial. The horizontal
forces are well-defined only if one {\it chooses} fixed reference
values for, say, the $F^{(i,1)}_3$'s. If these values are large
enough, as explained in Eq. (\ref{FGtransform}) and therebelow, all
the horizontal forces are positive irrespective of the $G_{i,j}$ values
within the bounds defined by inequality (\ref{Gineq}). Then the set
${\cal S}_p$ of allowed values of the parameters  $G_{i,j}$ is the 
intersection of the hyperplane ${\cal
S}_h$ defined by Eq. (\ref{period}) and the polyhedron defined by the
inequality  (\ref{Gineq}).

\subsubsection{The q-coordinates:\label{sec2.2.2}}

There is an alternative formulation of this problem in terms of  new
variables
\begin{eqnarray}
q_{i,j}=\frac{\sqrt{3}(G_{i,j}+F^{(i,j)}_2)}{2W_{i,j}}\,,
\label{varchange}
\end{eqnarray}
where $q_{i,j}$ is the fraction of $W_{i,j}$ that the $(i,j)$th
particle transmits to the layer below it towards the left, i.e., $F^
{(i,j)}_5= \frac{2}{\sqrt{3}}q_{i,j} W_{i,j}$ and
$F^{(i,j)}_6=\frac{2}{\sqrt{3}}(1-q_{i,j}) W_{i,j}$. The mapping
(\ref{varchange}) then reduces inequality (\ref{Gineq}) to $0 \leq
q_{i,j} \leq 1$. Clearly,  $W_{1,j_0}=W_{\mathrm{ext}}$ is the
external force applied on the top layer. For $i>1$, the $W_{i,j}$'s in
the successive layers are related via
\begin{eqnarray}
W_{i,j}=(1-q_{i-1,j-1})\,W_{i-1,j-1}+q_{i-1,j}\,W_{i-1,j}
\label{qprop}
\end{eqnarray}
i.e., $W_{i,j}$ is a function of $q_{k,l}$ for $k<i$.

From the formulation in terms of the $q$'s it may seem from
Eq. (\ref{qprop}) that in the hexagonal geometry of Fig. \ref{fig1},
one simply recovers the $q$-model \cite{qmodel}. There is however an
important subtlety to take notice of. In the $q$-model, the $q$'s
corresponding to different grains are usually uncorrelated. In our
case, the uniform measure on ${\cal E}$ implies, from
Eq. (\ref{varchange}), that
\begin{equation}
{\prod\limits_{i,j}}\,dG_{i,j}=\left[\frac{2}{\sqrt{3}}\right]^{\frac{N(N-1)}{2}}\,{\prod\limits_{i,j}}
dq_{i,j}\,W_{i,j}(q)\,,
\label{jac}
\end{equation}
i.e., due to the presence of the Jacobian on the right hand side of
Eq. (\ref{jac}), the uniform measure on ${\cal S}$ translates to a
non-uniform measure on the $\frac{N(N+1)}{2}$-dimensional unit cube
formed by the accessible values of the $q$'s.
 In the case of periodic boundary conditions, the constraint 
(\ref{period}) reduces to
\begin{eqnarray}
\sum_{j=1}^{p} W_{i,j}q_{i,j}=\sum_{j=1}^{i-1}
W_{i-1,j}q_{i-1,j}=\frac{1}{\sqrt{3}} W_{\mathrm{ext}}.
\label{qperiod}
\end{eqnarray}

\section{Numerical results \label{sec3}}

\subsection{Shape and scaling of the response \label{sec3.1}}
\begin{figure*}
\begin{center}
\begin{minipage}{0.3\linewidth}
\includegraphics[width=0.9\linewidth]{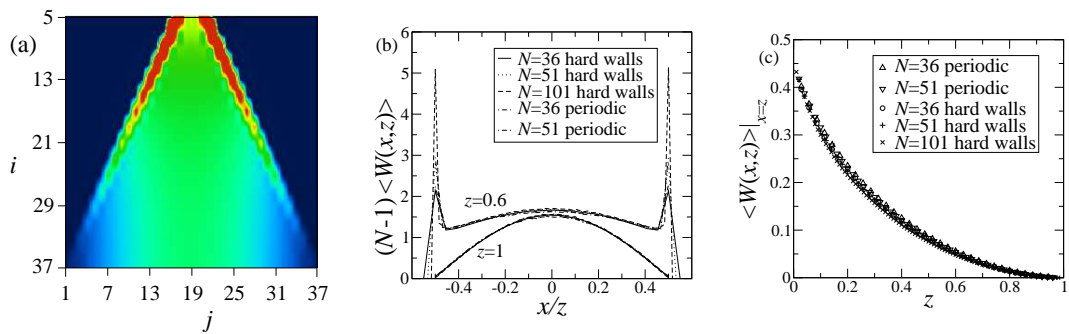}
\end{minipage}
\begin{minipage}{0.3\linewidth}
\includegraphics[width=0.9\linewidth]{profiles_new.eps}
\end{minipage}
\begin{minipage}{0.31\linewidth}
\includegraphics[width=0.9\linewidth]{scaling_new.eps}
\end{minipage}
\end{center}
\noindent \caption{Simulation results for $\langle W_{i,j}\rangle$:
(a) colour plot for $\langle W_{i,j}\rangle$ with hard walls;
$N=37$. The null vanishing forces appear in deep indigo, while the
largest are represented in dark red. The other values are represented
by a linear wavelength scale in between.(b)-(c) Behavior of  $\langle
W_{i,j}\rangle$ in reduced co-ordinates $x$ and $z$ for different
$N$-values and boundary conditions: (b) scaling of $\langle
W(x,z)\rangle$ with system size for $|x|<z$ at two $z$ values; (c)
data collapse for $\langle W(x,z)\rangle|_{x=z}$ for different $N$
values and boundary conditions. See text for further
details.\label{weightless}}
\end{figure*}

We now present the results for the $\langle W_{i,j}\rangle$'s
evaluated numerically by implementing detailed balance with respect to
the probability measure $d\mu=\frac{1}{\cal{N}}\prod\limits_{i,j}
dq_{i,j}\,W_{i,j}(q)\,$ on the
$\frac{N(N-1)}{2}$-dimensional unit cube in the $q$-space. At each step, a single $q$ is modified, and the change is
accepted with a Metropolis acceptance ratio. From
Eq. (\ref{qprop}), it is easy to see that the $\langle W_{i,j}\rangle$
values scale linearly with that of $W_{\mathrm{ext}}$. Thus, from now
on, we set $W_{\mathrm{ext}}=1$.

Our simulation results for $\langle W_{i,j}\rangle$ in the case of
hard walls with $F^{(i,1)}_3=\frac{2}{\sqrt{3}}W_{\mathrm{ext}}$ for
$N=37$ are plotted in Fig. \ref{weightless}(a), using the built-in
cubic interpolation function of Mathematica. Outside the triangle
shown in Fig. \ref{fig1}(b), $\langle W_{i,j}\rangle\equiv0$ appears
in deep indigo; the largest $\langle W_{i,j}\rangle$ value within the
triangle appears in dark red; and any other non-zero $\langle
W_{i,j}\rangle$ value is represented by a linear wavelength scale in
between. The thin green  regions that appear on the outer edge of the
triangle are artifacts of the interpolation.

We find for any system size $N$ that the $\langle W_{i,j}\rangle$
values display two symmetric peaks that lie precisely on the two
downward lattice directions emanating from the point of application of
the force $W_{\mathrm{ext}}$. The width of these peaks is a single
grain diameter and with depth their magnitudes decrease, while a
broader central peak  appears. At the very bottom layer ($i=N$) only
the central maximum for  $\langle W_{i,j}\rangle$ remains.

We further define $x=\frac{j-j_0}{N}$ and $\frac{j-j_0+1/2}{N}$
respectively for even and odd $i$, and $z=\frac{i}{N}$ [see
Fig. \ref{fig1}] in order to put the vertices of the triangle formed
by the locations of non-zero $\langle W_{i,j}\rangle$ values on
$(0,0), (-1/2,1)$ and $(1/2,1)\,\,\forall N$. The excellent data
collapse indicates that the $\langle W_{i,j}\rangle$ values for
$|x|<z$ scale with the inverse system size [Fig. \ref{weightless}(b);
we however show only two $z$ values], while the $\langle
W_{i,j}\rangle$ values for $|x|=z$ lie on the same curve for all
system sizes [Fig. \ref{weightless}(c)]. The data suggest that in the
thermodynamic limit $N\rightarrow\infty$, the {\it response field\/}
$\langle W(x,z)\rangle$ scales $\sim 1/N$ for $|x|<z$, but reaches a
{\it non-zero\/} limiting value on $|x|=z\,\,\forall z<1$. We thus
expect $\lim\limits_{N\rightarrow\infty}\langle
W(x,z)\rangle\big|_{|x|=z}>\langle
W(x,z)\rangle\big|_{|x|<z}\,\,\forall z<1$; or equivalently, a {\it
double-peaked response field at all depths $z<1$ in the thermodynamic
limit}.

\begin{figure*}[t]
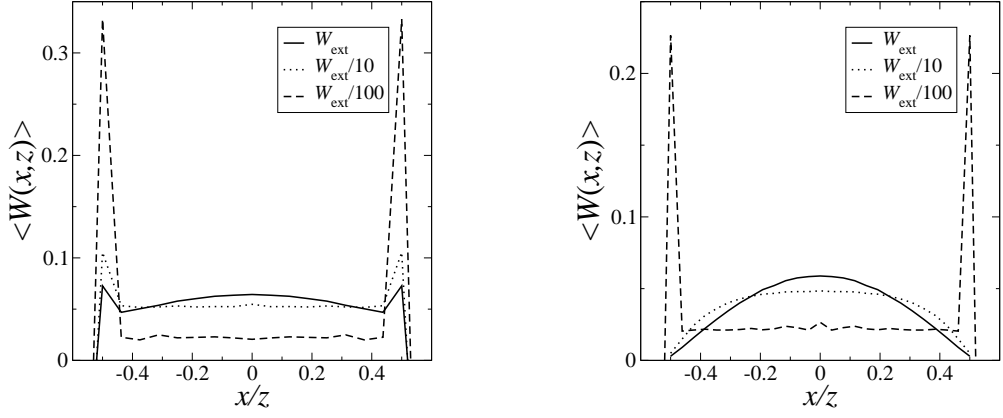

\begin{center}
\begin{minipage}{0.4\linewidth}
\includegraphics[width=0.9\linewidth]{side_f1}
\end{minipage}
\hspace{1cm}
\begin{minipage}{0.4\linewidth}
\includegraphics[width=0.9\linewidth]{side_f2}
\end{minipage}
\end{center}
\noindent \caption{Responses in reduced coordinates for systems with
three different values of $\frac{\sqrt3}{2}F^{i,1}_3$ at two different
depths, with periodic boundary conditions: (a) at $z=0.75$, (b) at
$z=1$.\label{side_f} }
\end{figure*}

The introduction of the additional constraint (\ref{qperiod}) that
correspond to periodic boundaries modifies neither the qualitative,
nor the scaling properties of the response function as can be observed
in Fig. \ref{weightless}(b)-(c).

Although at all places in this paper we consider each side force
$F^{(i,1)}_3=\frac{2}{\sqrt3}W_{\mathrm{ext}}$ so that the full range
of $q$-values are allowed for each $q_{i,j}$, in this paragraph, we
take a short digression to discuss what happens when one chooses
$F^{(i,1)}_3$'s to be smaller than $\frac{2}{\sqrt3}W_{\mathrm
ext}$. At the extreme limit when $F^{(i,1)}_3=0$, it is easy to see
that the only grains that correspond to non-zero $W_{i,j}$-values lie
exactly on the boundary. From there on, it is intuitively clear that
once the values of the $F^{(i,1)}_3$'s are pushed higher, the {\it
finite\/} range of allowed $q$-values would begin to transfer some of
the vertical forces into the bulk. A higher range of allowed values of
the $q$'s, caused by higher values of $F^{(i,1)}_3$'s, would thus
effectively result in smaller peaks on the boundary and
correspondingly, bigger fractions of the total vertical force within
the bulk.

 Fig. \ref{side_f}, where we plot the response at two different $z$
values ($z=0.75$ and $z=1$) for $N=51$ and
$\frac{\sqrt3}{2}F^{(i,1)}_3=W_{\mathrm{ext}}$, $W_{\mathrm{ext}}/10$
and $W_{\mathrm{ext}}/100$ respectively with periodic boundary
conditions, clearly supports such an intuitive picture.

\subsection{Linearity of response\label{sec3.2}}

\begin{figure*}[h]
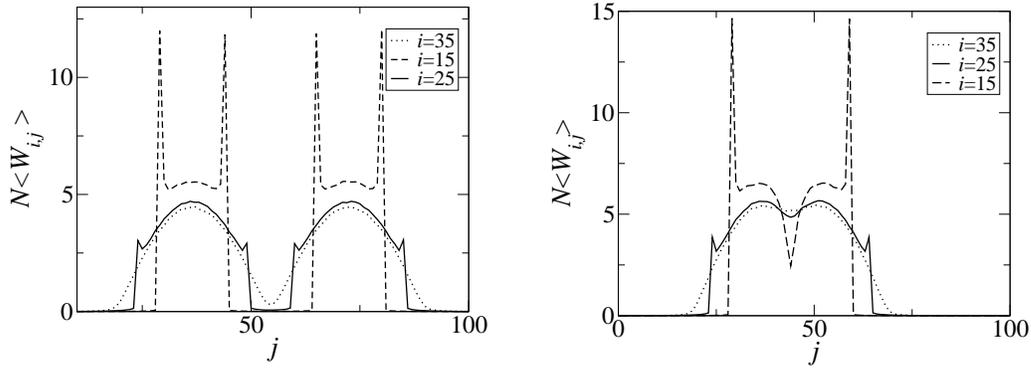

\begin{center}
\begin{minipage}{0.45\linewidth}
\includegraphics[width=0.9\linewidth]{sup1}
\end{minipage}
\begin{minipage}{0.45\linewidth}
\includegraphics[width=0.9\linewidth]{sup2}
\end{minipage}
\caption{Simulation results for the response $N\langle W_{i,j}\rangle$
due to two vertically downward forces of unit magnitude each applied
on the grains $(1,j_1)$ and $(1,j_2)$ for $N=35$ and $p=100$: (a)
$j_2-j_1=35$ (b) $j_2-j_1=15$. Three different values of $i$ are
displayed in each case. \label{superp}}
\end{center}
\end{figure*}

Since the values of the $\langle W_{i,j}\rangle$'s trivially scale
linearly with the magnitude of $W_{\mathrm{ext}}$, a natural question
is whether the response depends linearly on $\mathsf
F_{{\mathrm{ext}}}$ in general, i.e., whether the response to a
superposition of external forces is simply the superposition of
responses.

It might appear at first sight that the response is indeed linear ---
after all, the heart of the problem is the system of linear equations
${\mathbf A}\cdot{\mathsf F}={\mathsf F}_{\mathrm{ext}}$. In fact,
with the notations of Sec. \ref{sec2.1}, one might argue that since
${\mathsf F}^{(0)}$ is a linear function of ${\mathsf
F}_{\mathrm{ext}}$, while $Ker A$ is fixed for a given contact
geometry, the averages over the affine space ${\mathsf F}^{(0)}+Ker A$
should depend linearly on ${\mathsf F}_{\mathrm{ext}}$. Notice however
that the set ${\cal E}$ corresponds {\it only\/} to $F_{k} \geq 0$,
and the shape and volume of ${\cal E}$ in general depend on ${\mathsf
F}_{\mathrm{ext}}$ in a complex way, and thus the response needs not
be linear.

To illustrate this point,  we apply two vertically downward forces of
unit magnitude on the  $(1,j_1)$th and $(1,j_2)$th grain for $N=35$
and $p=100$. The $|j_2-j_1|>N$ case is a trivial situation, since in
this case the two triangular regions of non-zero $W_{i,j}$'s do not
overlap with each other, and the system is simply the superposition of
two  stochastically independent subsystems. In other words, the
response observed for the $|j_2-j_1|>N$ is simply a superposition of
the responses of two individual responses [see
Fig. \ref{superp}(a)]. For $|j_2-j_1|\leq N$ however, the two
triangular regions do overlap and the subsystems interact; as shown in
Fig. \ref{superp}(b), the fact that there exists a central {\it
minimum\/} for the response at $i=15$ is a clear demonstration
that the linearity of the responses does not hold.

\subsection{q-distributions and correlations\label{sec3.3}}

\begin{figure*}
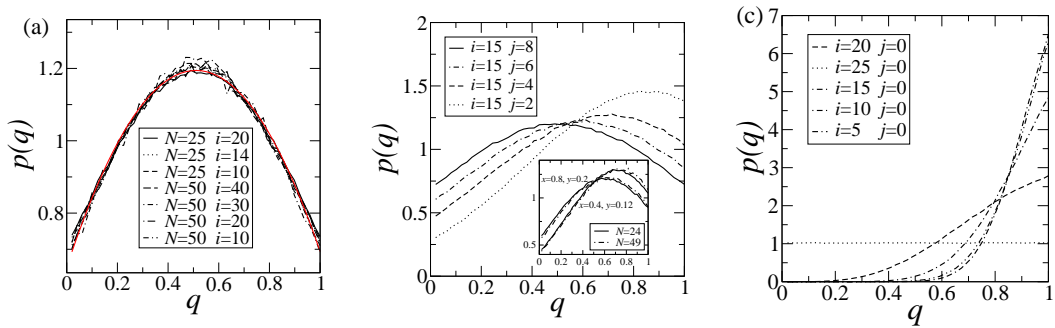

\begin{center}
\begin{minipage}{0.3\linewidth}
\includegraphics[width=0.9\linewidth]{qdist3}
\end{minipage}
\begin{minipage}{0.3\linewidth}
\includegraphics[width=0.9\linewidth]{qdist1}
\end{minipage}
\begin{minipage}{0.3\linewidth}
\includegraphics[width=0.9\linewidth]{qdist2}
\end{minipage}
\caption{$p(q_{i,j})$ for (a) $j=i/2$ for $N=25$ and $50$, with a
quadratic fit in red (b) for $N=25$ at $i=15$ for four different $j$
values [inset: self-similarity of $p(q_{i,j})$ for $N=25$ and $N=50$ at
$(x,z)= (0.12,0.4)$ and$(x,z)=(0.2, 0.8)$], (c) for $N=25$ on the left
boundary $j=0$ at different heights $i$. \label{qdist}}
\end{center}
\end{figure*}

We have just seen that the qualitative behaviour of our model is very
different from that of the $q$-model. It thus seems reasonable that we
study the difference between these two models at the level of
individual grains.

Our starting hypothesis of equal probability for all mechanically
stable force configurations implies that the joint probability
distribution for the $q_{i,j}$'s is given by $\frac{1}{\cal
N}\prod_{ij}W_{i,j}(q)$, where ${\cal N}$ is the normalization
constant. This distribution does not factorize into a product of terms
that depend on single $q_{i,j}$'s, and as a result, the $q_{i,j}$'s
associated with different grains are correlated with each other. Below
we numerically investigate the properties of the induced probability
distributions $p(q_{i,j})$ for single $q_{i,j}$'s, as well as the
correlations between different $q_{i,j}$'s throughout the system.

Since individual $W_{i,j}(q)$'s are independent of $q_{N,k}$'s for
$k=1,\ldots,N$, each of the $q_{N,k}$'s is uncorrelated with all other
other $q$'s in the system, and is also uniformly distributed within
the interval $[0,1]$. However, higher up in the packing, the
$q$-distributions start to show a single maximum. For an odd $i$ and
$j_0=\frac{i+1}{2}$, i.e., at the centre of the layer, $p(q_{i,j_0})$
exhibits a single maximum precisely at $q_{i,j_0}=0.5$ due to symmetry
reasons. Furthermore, $p(q_{i,j_0})$ is independent of the precise
value of $i<N$ as well as the system size, and it is well-fitted by a
quadratic polynomial [see Fig. 6(a)]. For a given layer $i$, the
larger $|j-j_0|$ is, the more the location of the single maximum of
$p(q_{i,j})$ shifts away (symmetrically) from $q_{i,j}=0.5$: for
$j-j_0>0$, this maximum occurs at $q_{i,j}>0.5$ and for $j-j_0<0$,
this maximum occurs at $q_{i,j}<0.5$ [Fig. 6(b)]. Finally, exactly on
the boundary the maximum of $p(q_{i,j})$ occurs at $q=0$ (left
boundary) or at $q=1$ (right boundary) [Fig. 6(c)].

Interestingly, the self-similarity that we observed for the response
behaviour also seems to be valid for $p(q_{i,j})$ in the bulk [see the
inset of Fig. 6(b)]. This stands at a stark contrast to the physical
behaviour of the $q$-model, as the self-similarity implies that in our
model, the forces in a granular packing do {\it not\/} propagate from
top down, instead
they depend on the whole extent of the system.

In Fig. 7, we show results for the correlations between $q_{i_c,j_c}$
and $q_{i,j}$ for several values of $i$ and $j$ for $N=50$,
$(i_c,j_c)=(25,12)$ and $(i_c,j_c)=(47,10)$. It appears from the plots
that the correlations between the $q$'s at different locations are
very weak.

\begin{figure*}
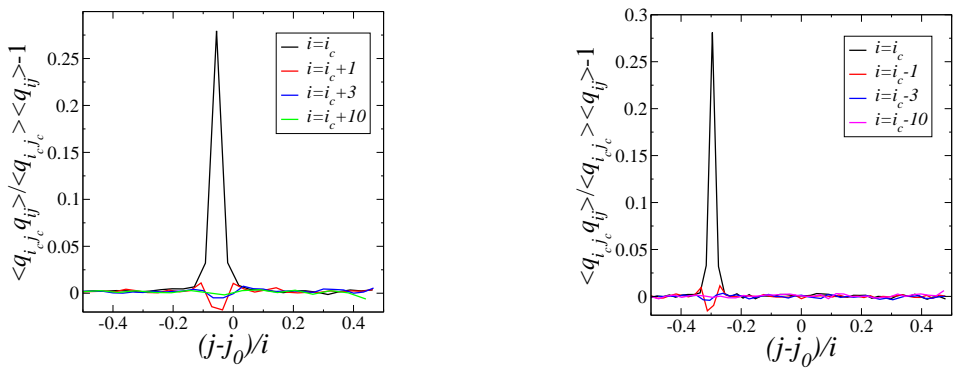

\begin{center}
\begin{minipage}{0.4\linewidth}
\includegraphics[width=0.8\linewidth]{corr1}
\end{minipage}
\hspace{1cm}
\begin{minipage}{0.4\linewidth}
\includegraphics[width=0.8\linewidth]{corr2}
\end{minipage}
\caption{Correlations between $q_{i_c,j_c}$ and $q_{i,j}$ at various
values of $i$ for $N=50$, (a) $(i_c,j_c)=(25,12)$ and (b)
$(i_c,j_c)=(47,10)$.\label{corr}}
\end{center}
\end{figure*}

\section{Exact Results for $\langle W_{i,j}\rangle$: Summary
\label{sec4new}} 

Having presented the simulation results above, from this section
onwards we concentrate on the exact (theoretical) results. It turns
out that in this model not only the $\langle W_{i,j}\rangle$ values,
but also all the higher moments and correlations of the $W_{i,j}$'s
can be expressed exactly in terms of integers defined by  simple recursion relations. The full calculational details
for the exact results for $\langle W_{i,j}\rangle$ will be presented in
Sec. \ref{sec4}. The details of the procedure are slightly involved, and
therefore we provide a layout summary of results and methods in this section.

Our main result are exact expressions for all moments of $W_{i,j}$ for
any system size $N$ in terms of integers defined by recursive
relations. For instance, the zero-th moment or normalization constant
${\cal N}$
for a packing of $N$ layers reads

\begin{eqnarray}
{\cal
N}=\frac{1}{\left[\frac{(N-1)N}{2}\right]!}\,\,\sum_{j_1,\ldots,j_{N-2}}
\Lambda^{(N-2)}_{j_1,\ldots,j_{N-2}}\,.
\end{eqnarray}

where $\Lambda^{(N-2)}_{j_1,\ldots,j_{N-2}}$ are integers indexed by
$\{j_k\}_{1\leq k \leq N-2}$, with $1\leq j_1 \leq 2$ and $1\leq j_k
\leq k+j_{k-1}$. They are given by the following reccurrence
relations

\begin{eqnarray}
\Lambda^{(2)}_{j_1j_2}&=&\delta_{j_2,1}\nonumber\\
\Lambda^{(p+1)}_{j_1,\ldots,j_{p+1}}&=&\delta_{j_{p+1},1}\sum_{k_1,\ldots,k_p}\Lambda^{(p)}_{k_1,\ldots,k_{p}}\prod_{l=1}^{p}\Theta(j_{l+1}-k_l)\,.
\label{lambda_form_sum}
\end{eqnarray}

with $\Theta$ the discrete Heaviside (step) function.

Any higher moment of $W_{i,j}$ can be expressed as
$\frac{1}{\cal{N}}\sum_{j_1,\ldots,j_{N-2}}\bar{\Lambda}^{(N-2)}_{j_1,\ldots,j_{N-2}}$,
where $\bar{\Lambda}^{(N-2)}_{j_1,\ldots,j_{N-2}}$ are integers
obtained through slight modifications of the reccurence formulas
(\ref{lambda_form_sum}), described in full detail in sections \ref{sec4.3.2},
\ref{sec4.3.3}, and \ref{sec4.4}. These results establish a
non-trivial equivalence between the model studied here and a discrete
combinatorial problem defined in Sec.\ref{sec4.6}.

The crux of calculating any moment of $W_{ij}$ lies in the transformation of the integration measure
$\prod\limits_{i,j}dG_{i,j}$ [such as in Eq. (\ref{waverage})]. In
Eq. (\ref{jac}), $\prod\limits_{i,j}dG_{i,j}$ has been rewritten in
terms of the $q_{i,j}$'s. With another change of variables from
$q_{i,j}$'s to $W_{i,j}$'s, the integration measure can be further
expressed as
\begin{eqnarray}
{\prod\limits_{i,j}}\,dG_{i,j}=\left[\frac{2}{\sqrt{3}}\right]^{\frac{N(N-1)}{2}}\,{\prod\limits_{i,j}}
dq_{i,j}\,W_{i,j}(q)=\left[\frac{2}{\sqrt{3}}\right]^{\frac{N(N-1)}{2}}{\prod\limits_{i,j}}\,dW_{i,j}\,.
\label{e13new}
\end{eqnarray}

While such a variable change simplifies the integrands, the mapped volume
${\cal S}$ (see the last paragraph of Sec. \ref{sec2.1}) is however not an $N$-dimensional square anymore, but again a
polygone.
Despite this complication, one can build up
a {\it recursive structure for the integrations over $W_{k,l}$'s},
which allows to calculate recursively the integrals over ${\cal S}$.

\begin{figure*}
\begin{center}
\begin{minipage}{0.40\linewidth}
\includegraphics[width=0.95\linewidth]{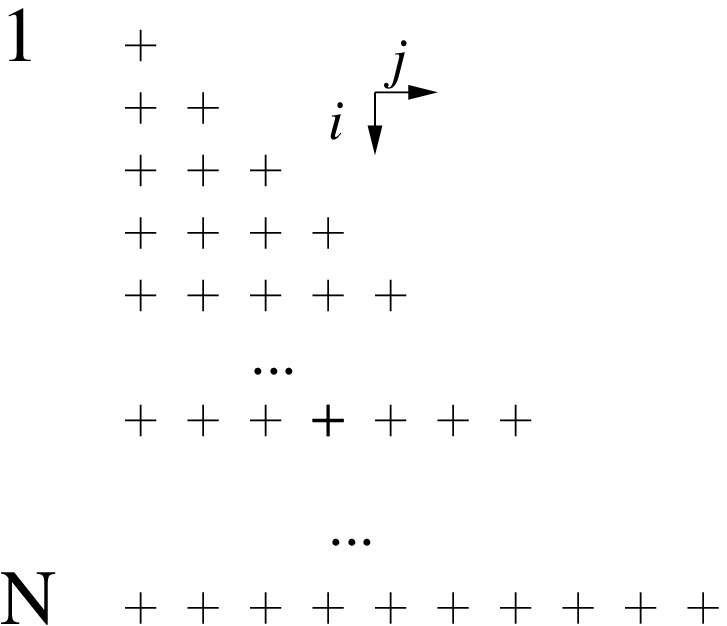}
\end{minipage}
\begin{minipage}{0.45\linewidth}
\includegraphics[width=0.95\linewidth]{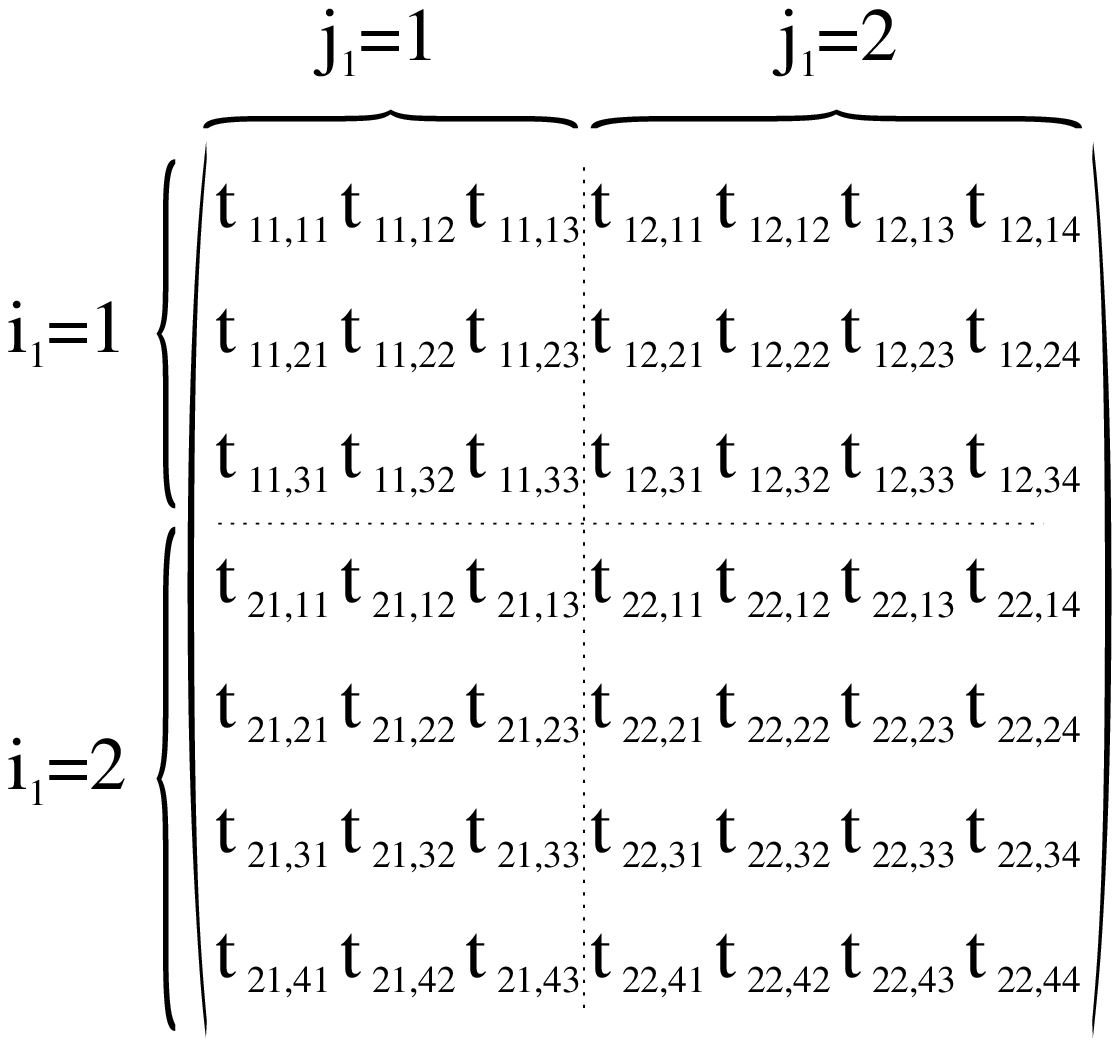}
\end{minipage}
\end{center}
\caption{(a) Schematic view of the sublattice we consider; (b)
illustration of the ``nested''  notation used for the matrix
elements\label{mesh1} for $p=2$.}
\end{figure*}

Before we start to calculate any of the integrals, for convenience we
first distort the triangle in Fig. \ref{fig1}(b) to that of
Fig. \ref{mesh1}(a). We choose $j_0=1$, and relabel the grains in the
$i$-th row as $(i,j)$ with $1\leq j\leq i$. Starting at
$(k,l)=(N,N)$, we then execute the integrals over
$W_{k,l}$ in the decreasing order of both $k$ and $l$; i.e., for a given
value of $k$, we integrate over $W_{k,l}$ sequentially for decreasing
$l$, we then decrease $k$ by one, and continue the integration process
over $W_{k-1,l}$ sequentially for decreasing $l$, and so on. The
integration results over the successive layers are expressed in a
``hierarchical nested matrix form'' in a recursive manner.

To illustrate the recursive formulation of the ``hierarchical nested
matrix form'', let us consider the calculation of ${\cal N}$. In the
expression of ${\cal
N}=\displaystyle{\int_{{\cal S}}{\prod\limits_{k,l}}\,dW_{k,l}}$, evaluating the
integrations over $W_{N,N-1}$ through $W_{N,1}$, and $W_{N-1,N-2}$
through $W_{N-1,1}$, we find that the result is a polynomial in
$W_{N-2,k}$ that  can be written as a
matrix product
$J_{N-2}=\displaystyle{\int{\prod^{N-2}_{l=1}}\,dW_{N-1,l}{\prod^{N-1}_{k=1}}\,dW_{N,k}
}=\left[L^{(1)}\right]^{\mathrm{T}}\,\prod_{l=1}^{N-2}
{t}^{N-1,l}\,R^{(1)}$. Here, $L^{(1)}$ and $R^{(1)}$ are respectively
$1\times2$ and $2\times1$ matrices, 
and the elements of the $2\times2$ matrix $t^{N-1,l}$  depend only on $W_{N-2,l}$. Thereafter, when $J_{N-2}$ is integrated over $W_{N-2,N-3}$
through $W_{N-1,1}$, the corresponding result
$J_{N-3}=\displaystyle{\int{\prod^{N-3}_{l=1}}\,dW_{N-2,l}}\,J_{N-2}$
yields again a polynomial expressible in a similar matrix product form, i.e.,
$J_{N-3}=\left[L^{(2)}\right]^{\mathrm{T}}\,\prod_{l=1}^{N-3}
{t}^{N-2,l}\,R^{(2)}$.
The {\it crucial point to note however is that the matrices $L^{(2)}$,
$t^{N-2,l}$ and $R^{(2)}$ can be constructed by simply unfolding each
respective element of $L^{(1)}$, $t^{N-1,l}$ and $R^{(1)}$ as
matrices} [see Fig. \ref{mesh1}(b) for the elements of $t^{N-2,l}$,
which are now indexed by $i_1j_1,i_2j_2$]. After the integration of
$J_{N-3}$ over the $W_{N-3,l}$'s, each of the elements of $L^{(2)}$,
$t^{N-2,l}$ and $R^{(2)}$ further unfold into matrices and so on. This
process continues over further and further integrations generating the
``hierarchical nested matrix form''.  The elements of
$L^{(k-1)}$, $t^{N-k+1,l}$ and $R^{(k-1)}$ are related to those of
$L^{(k)}$, $t^{N-k,l}$ and $R^{(k)}$ in a recursive
manner.

In Sec. \ref{sec4}, we develop the full details of this recursive
integration scheme and calculate $\langle W_{i,j}\rangle$ exactly. In
Sec. \ref{prelim}, we establish some priliminary
formulas, which we use over and over again during the course of the exact
calculation. In Secs. \ref{norm_sect} and \ref{expect} respectively,
we evaluate ${\cal N}$ and
$\displaystyle{\int{\prod\limits_{k,l}}\,dW_{k,l}\,\,W_{i,j}}$. How to
proceed with the calculations of the higher moments and correlations
of $W_{i,j}$'s is discussed in Sec. \ref{sec4.4}.

\section{Exact Results for $\langle W_{i,j}\rangle$:
Details\label{sec4}}

\subsection{Preliminaries \label{prelim}}



As described in Eqs. (\ref{varchange}-\ref{jac}), a full description
of a given realization of the contact force configurations is given by
the variables $\{q_{i,j}\}_{i=1\ldots N-1,j=1\ldots i}$ with $0\leq
q_{i,j}\leq1$. The unnormalized joint probability distribution on these
$q$-variables is $\prod_{i=1}^{N-1}\prod_{j=1}^{i} W_{i,j}$.  Here,
$W_{1,1}=W_{\mathrm{ext}}=1$, and for $i>1$, the successive
$W_{i,j}$'s are given by
\begin{eqnarray}
W_{i,1}=q_{i-1,1} W_{i-1,1} \nonumber \\
W_{i,j}=(1-q_{i-1,j-1})\,W_{i-1,j-1}+q_{i-1,j}\,W_{i-1,j} \qquad
\mathrm{for} \quad 1<j<i\,\,\mbox{and}\nonumber\\
W_{i,i}=(1-q_{i-1,i-1}) W_{i-1,i-1}\,.
\label{qprop_d}
\end{eqnarray}

We calculate the moments of $W_{i,j}(q)$ on this joint
$q$-distribution by changing variables from $\{q_{i,j}\}_{i=1\ldots
N-1,j=1\ldots i}$ to $\{W_{i,j}\}_{i=2\ldots N,j=\ldots i-1}$. This is
achieved by rewriting Eq. (\ref{qprop_d}) as
\begin{eqnarray}
q_{i,j}=\frac{1}{W_{i,j}} \left[\sum_{k=1}^{j}
W_{i+1,k}-\sum_{k=1}^{j-1} W_{i,k}\right]\,.
\label{qWchange}
\end{eqnarray}
Notice from Fig. 1(b) that since all the forces on the triangle from
the left and the right are horizontal, $\sum_{k=1}^{i}
W_{i,k}=1\,\forall i$. This implies that on the $i$-th layer there are
only $(i-1)$ unconstrained $W_{i,j}$'s. We choose them to be $W_{i,j}$
for $1\leq j\leq i-1$. Then $W_{i,i}=1-\sum_{k=1}^{i-1} W_{i,k}$.

The advantage associated with the change of variables from
$\{q_{i,j}\}_{i=1,\ldots, N-1,j=1,\ldots,i}$ to
$\{W_{i,j}\}_{i=2,\ldots,N,j=1,\ldots,i-1}$ is that
$\prod_{i=1}^{N-1}\prod_{j=1}^{i}dq_{i,j}=\left[\prod_{k=1}^{N-1}\prod_{l=1}^{k}W_{k,l}\right]^{-1}\prod_{i=2}^{N}\prod_{j=1}^{i-1}dW_{i,j}$,
so that [from Eq. (\ref{jac})]
$\prod_{ij}dG_{i,j}=\prod_{ij}dW_{i,j}$, i.e., the $W_{i,j}$'s are
uniformly distributed. The difficulty of this formulation, however,
is that the volume the $W_{i,j}$'s span is not a cube
anymore. Instead, the volume spanned is a polygon given by the
inequalities
\begin{eqnarray}
a_{i,j} \leq W_{i,j} \leq b_{i,j},
\label{abdef}
\end{eqnarray}
with
\begin{eqnarray}
a_{i,j}=\sum_{k=1}^{j-1} (W_{i-1,k}-W_{i,k})\nonumber \\
b_{i,j}=W_{i-1,j}+ \sum_{k=1}^{j-1} (W_{i-1,k}-W_{i,k})
\end{eqnarray}
The $a_{i,j}$'s and the $b_{i,j}$'s are related by the following
simple relations:
\begin{eqnarray}
b_{i,j}- a_{i,j} & = & W_{i-1,j} \label{abprop1}\\ a_{i,j} & = &
b_{i,j-1}-W_{i,j-1} \label{abprop2}\\ b_{i,1} & = & W_{i-1,1}
\label{abprop3}\\ a_{i,1} & = & 0\,.  \label{abprop4}
\end{eqnarray}

Finally, the following integral is the centrepiece of all our
calculations:
\begin{eqnarray}
I_{mn}(a,b)=\int_{a}^{b} x^m (b-x)^n dx=\sum_{k=0}^{m}
\alpha_{mn}^{m-k}\,(b-a)^{n+1+k}\,a^{m-k}\,,
\label{Idev}
\end{eqnarray}
where
\begin{eqnarray}
\alpha_{mn}^{m-k}=\frac{m!\,n!}{(n+k+1)!\,(m-k)!}\,.
\label{alpha}
\end{eqnarray}

\subsection{Evaluation of the Normalization constant ${\cal N}$
\label{norm_sect}} 

In the $W_{i,j}$-space, the average of a quantity $h$ is given by
$\langle h\rangle=\left[\int\prod_{ij}dW_{i,j}h\right]/{\cal N}$,
where ${\cal N}=\int\prod_{ij}dW_{i,j}$ is the normalization constant.
In Sec. \ref{norm_sect}, we evaluate ${\cal N}$ by performing the
integrations over $\prod_{ij}W_{i,j}$ layer by layer bottom up; i.e.,
we start at $i=N$ and decrease $i$ until we reach the top layer where
$i=1$. At each layer, the integrations are carried out one by one in
the direction of decreasing $j$, from $j=i-1$ to $j=1$. The integral
$J_{N-p}=\int\prod_{i=N-p+1}^N\prod_{j=1,\ldots,i-1}dW_{i,j}$ is the
unnormalized induced probability density of the $W_{i,j}$'s for the
$(N-p)$ top layers. In this notation, ${\cal N}=J_1$.

Our exact calculations are made possible due to the particular forms
of the bounds $a_{i,j}$'s and $b_{i,j}$'s, as they introduce a
recursive structure. We will evaluate $J_{N-1}$ and $J_{N-2}$
explicitly, after which we will prove the general recurrence relation
for $J_{N-p}$ by induction. In fact, we will show that $J_{N-p}$ can
be written as a matrix product. The matrices entering this product are
given by recursive relations, which we will call the ``fundamental
relations'', as they are the building blocks of the calculation of all
moments of $W_{i,j}$.

\subsubsection{Evaluation of $J_{N-1}$ and $J_{N-2}$: \label{first2}}

By definition, $J_{N-1}= \displaystyle{\int \prod_{j=1}^{N-1}
dW_{N,j}}$.  We integrate the $W_{N,j}$'s one by one in the direction
of decreasing $j$'s. Using
\begin{eqnarray}
\int_{a_{N,j-1}}^{b_{N,j-1}}
dW_{N,j-1}=b_{N,j-1}-a_{N,j-1}=W_{N-1,N-1}\,,
\label{e22}
\end{eqnarray}
we have
\begin{eqnarray}
J_{N-1}=\int \prod_{j=1}^{N-1}dW_{N,j}=\prod_{j=1}^{N-1}W_{N-1,j}\,.
\nonumber
\end{eqnarray}

In order to obtain $J_{N-2}$, we have to integrate $J_{N-1}$ w.r.t.
$W_{N-1,j}$ for $1\leq j \leq N-2$. Since the bounds $a_{i,j}$ and
$b_{i,j}$ depend only on $W_{i,l}$ and $W_{i-1,l}$ for $l<j$,
$J_{N-2}$ can be expressed in a nested form:
\begin{eqnarray}
\hspace{-1cm}J_{N-2}&=&\int
\left[\prod_{j=1}^{N-2}dW_{N-1,j}\right]\,J_{N-1} \nonumber\\
&=&\!\!\int_{a_{N\!-\!1,1}}^{b_{N\!-\!1,1}}\!\!dW_{N\!-\!1,1}
W_{N\!-\!1,1} \ldots
\int_{a_{N\!-\!1,N\!-\!2}}^{b_{N\!-\!1,N\!-\!2}}\!\!dW_{N\!-\!1,N\!-\!2}\,W_{N\!-\!1,N\!-\!2}\,W_{N\!-\!1,N\!-\!1}\,,
\label{e23}
\end{eqnarray}
and thereafter it can be evaluated iteratively. Having defined
$J^{N-2,N-1}=W_{N-1,N-1}=1-\sum_{k=1}^{N-2} W_{N-1,k}$, we have
\begin{eqnarray}
J^{N-2,k}=\int_{a_{N-1,k}}^{b_{N-1,k}}
dW_{N-1,k}\,W_{N-1,k}\,J^{N-2,k+1}
\label{it1}
\end{eqnarray}
for $k>1$, i.e., $J_{N-2}=J^{N-2,1}$. We now show that $\forall
\,k\geq1$,  $J^{N-2,k}$ is of the form
$\gamma_{1}^{N-1,k-1}(b_{N-1,k-1}-W_{N-1,k-1})+\gamma_{2}^{N-1,k-1}$,
where the  vectors
$\gamma^{N-1,k-1}=\left[\begin{array}{c}\gamma_{1}^{N-1,k-1}\\\gamma_{2}^{N-1,k-1}\end{array}\right]$
and
$\gamma^{N-1,k}=\left[\begin{array}{c}\gamma_{1}^{N-1,k}\\\gamma_{2}^{N-1,k}\end{array}\right]$
are related by means of a matrix relation of the form
${\gamma}^{N-1,k-1}={t}^{N-1,k}\,{\gamma}^{N-1,k}$.

For $k=N-1$, the proposed form of $J^{N-2,k}$ is easily checked by
using $\sum_{k=0}^{N-2} W_{N-2,k}=1$ and Eq. (\ref{abdef}):
\begin{eqnarray}
\hspace{-1cm}J^{N-2,N-1}&=&1-\sum_{k=1}^{N-2}
W_{N-1,k}=\sum_{k=1}^{N-2}\left[W_{N-2,k}-W_{N-1,k}\right]=b_{N-1,N-2}-W_{N-1,N-2}\nonumber
\\
&=&\gamma_{1}^{N-1,N-2}\left[b_{N-1,N-2}-W_{N-1,N-2}\right]+\gamma_{2}^{N-1,N-2}\,,
\label{e25}
\end{eqnarray}
with $\gamma_{1}^{N-1,N-2}=1 $ and $\gamma_{2}^{N-1,N-2}=0$. This
means that $\gamma^{N-1,N-2}=\left[\begin{array}{c} 1\\0 \end{array}
\right]\equiv R^{(1)}$.

Let us now assume that the form
$J^{N-2,k+1}=\gamma_{1}^{N-1,k}(b_{N-1,k}-W_{N-1,k})+\gamma_{2}^{N-1,k}$
holds for a given $k$, $1<k<N-2$. Then from Eqs. (\ref{Idev}) and
(\ref{it1}), we have
\begin{eqnarray}
\hspace{-1.25cm}J^{N-2,k}=\gamma_{1}^{N-1,k}I_{11}(a_{N-1,k},b_{N-1,k})+\gamma_{2}^{N-1,k}
I_{10}(a_{N-1,k},b_{N-1,k})\nonumber\\ =\left[\gamma_{1}^{N-1,k}
\alpha_{11}^{1}(b_{N-1,k}-a_{N-1,k})^2+\gamma_{2}^{N-1,k}\alpha_{10}^{1}(b_{N-1,k}-a_{N-1,k})
\right] a_{N-1,k}\nonumber\\\hspace{4mm}+\left[\gamma_{1}^{N-1,k}
\alpha_{11}^{0}(b_{N-1,k}-a_{N-1,k})^3+\gamma_{2}^{N-1,k}\alpha_{10}^{0}(b_{N-1,k}-a_{N-1,k})^2\right]\,.
\label{e26}
\end{eqnarray}
Thereafter, using Eqs. (\ref{abprop1}) and (\ref{abprop2}),
$J^{N-2,k}$ can be rewritten as
\begin{eqnarray}
J^{N-2,k}=\gamma_{1}^{N-1,k-1}(b_{N-1,k-1}-W_{N-1,k-1})+\gamma_{2}^{N-1,k-1}\,,
\label{label21res}
\end{eqnarray}
with
\begin{eqnarray}
\gamma_{1}^{N-1,k-1} & = & \gamma_{1}^{N-1,k}
\alpha_{11}^{1}W_{N-2,k}^2+\gamma_{2}^{N-1,k}\alpha_{10}^{1}W_{N-2,k}\nonumber
\\ \gamma_{2}^{N-1,k-1} & = & \gamma_{1}^{N-1,k}
\alpha_{11}^{0}W_{N-2,k}^3+\gamma_{2}^{N-1,k}\alpha_{10}^{0}W_{N-2,k}^2\,.
\end{eqnarray}
In other words, we have the matrix relation
${\gamma}^{N-1,k-1}={t}^{N-1,k} \,{\gamma}^{N-1,k}$, where
${t}^{N-1,k}$ is a $2\times2$ matrix with elements
\begin{eqnarray}
t^{N-1,k}_{i_1j_1}=\alpha_{1,2-j_1}^{2-i_1}\,W^{2+i_1-j_1}_{N-2,k}\,.
\label{t2}
\end{eqnarray}

Thus, by means of the induction procedure via
Eqs. (\ref{e26}-\ref{t2}), we have demonstrated that
$J^{N-2,2}=\prod_{k=2}^{N-2}{t}^{N-1,k}\,R^{(1)}$.  Finally, in the
last integral (over $W_{N-1,1}$) in Eq. (\ref{e23}), the lower limit
$a_{N-1,1}=0$, and it is easily seen that
\begin{eqnarray}
\hspace{-2cm}J_{N-2}=\int_{a_{N-1,1}}^{b_{N-1,1}}\!\!dW_{N-1,1}\,W_{N-1,1}J^{N-2,2}=\left[L^{(1)}\right]^{\mathrm{T}}J^{N-2,1}=\left[L^{(1)}\right]^{\mathrm{T}}\,\prod_{k=1}^{N-2}
{t}^{N-1,k}\,R^{(1)},
\end{eqnarray}
where $\left[L^{(1)}\right]^{\mathrm{T}}$ is the row vector $[0\,\,1]$.

\subsubsection{Expression of $J_{N-p}$ by induction:}

We now derive the general formula for $J_{N-p}$ by induction. The
postulate is that the induced probability density of the $N-p$ top
layers can be written as
\begin{eqnarray}
J_{N-p}=\left[L^{(p-1)}\right]^{\mathrm{T}}
\prod_{k=1}^{N-p}{t}^{N-(p-1),k}\,R^{(p-1)}\,,
\label{e31}
\end{eqnarray}
where ${ t}^{N-(p-1),k}$, expressed in ``hierarchical nested matrix
form'', has elements
\begin{eqnarray}
{t}^{N-(p-1),k}_{i_1j_1,i_2j_2,\ldots,i_{p-1}
j_{p-1}}=\beta_{i_1j_1,i_2j_2,\ldots,i_{p-1} j_{p-1}}
W^{p+i_{p-1}-j_{p-1}}_{N-p,k}
\label{tform}
\end{eqnarray}
with
\begin{eqnarray}
\beta_{i_1j_1,\ldots,i_{p-1} j_{p-1}}=\alpha_{1,2-j_1}^{2-i_1}
\prod_{l=2}^{p-1}
\alpha_{l+i_{l-1}-j_{l-1},l+j_{l-1}-j_{l}}^{l+i_{l-1}-i_{l}}\Theta(i_l-j_{l-1})\,.
\label{beta}
\end{eqnarray}
Here $\Theta(k)$ is the discrete Heavyside Step-function, i.e.,
$\Theta(k)=1$ if $k\geq 0$, else $\Theta(k)=0$.  An alternative
recursive formulation for the $\beta$'s is given by
\begin{eqnarray}
\beta_{i_1j_1}& = &\alpha_{1,2-j_1}^{2-i_1}\qquad\mbox{and}\nonumber\\
\beta_{i_1j_1,\ldots,i_{p} j_{p}} & = & \beta_{i_1j_1,\ldots,i_{p-1}
j_{p-1}}
\,\alpha_{p+i_{p-1}-j_{p-1},p+j_{p-1}-j_{p}}^{p+i_{p-1}-i_{p}}\Theta(i_p-j_{p-1})\quad
\forall p>1.
\label{usual}
\end{eqnarray}
The expressions (\ref{tform}-\ref{usual}) will constantly be referred
to in all the following calculations, and we call them the
``fundamental relations''.

The vectors $L^{(p-1)}$ and $R^{(p-1)}$, respectively, are
\begin{eqnarray}
L_{i_1,\ldots,i_{p-1}}^{(p-1)}& =
&\prod_{l=1}^{p-1}\delta_{i_l,\frac{l(l+1)}{2}+1}\qquad\mbox{and}\nonumber\\
R_{i_1,\ldots,i_{p-1}}^{(p-1)}  & = &\delta_{j_{p-1},1}
\sum_{j_1\ldots j_{p-2}}\beta_{i_1j_1,\ldots,i_{p-2}
j_{p-2}}R_{j_1,\ldots,j_{p-2}}^{(p-2)}\,.
\label{LR}
\end{eqnarray}

``Hierarchical nested matrix form'' means that elements are referenced
through nested blocks. For example, the element referenced by
$i_1j_1,i_2j_2,\ldots,i_{p-1}j_{p-1}$ is the $i_{p-1}j_{p-1}$-th
sub-block of the block $i_1j_1,i_2j_2,\ldots,i_{p-2} j_{p-2}$, which
itself is the $(i_{p-1}j_{p-1})$-th sub-block of the block
$i_1j_1,i_2j_2,\ldots,i_{p-3}j_{p-3}$, and so on [see
Fig. \ref{mesh1}(b) for an illustration in the case $p=2$].  The
indices vary between the following bounds:
\begin{eqnarray}
1\leq i_1\leq 2, \qquad\qquad 1\leq j_1\leq 2\nonumber\\ 1\leq i_2\leq
2+i_1, \qquad 1\leq j_2\leq 2+j_1\nonumber\\
\hspace{1cm}\vdots\nonumber\\ 1\leq i_l\leq l+i_{l-1}, \qquad1\leq
j_l\leq l+j_{l-1}\label{ind_usual}\nonumber \\
\hspace{1cm}\vdots\nonumber\\ 1\leq i_{p-1}\leq p-1+i_{p-2}, \qquad
1\leq j_{p-1}\leq p-1+j_{p-2}\,.
\label{e36}
\end{eqnarray}
In our induction procedure, we assume the above forms
(\ref{e31}-\ref{e36}) for a given $p$, and then show that these
relations also hold when $p$ is replaced by $p+1$. The two step route
for this induction procedure that we follow is the same as the one we
followed to evaluate $J_{N-2}$.

Step (i): We first write
\begin{eqnarray}
J_{N-p-1}=\int \left[\prod_{k'=1}^{N-p-1}dW_{N-p,k'}\right]
\left[L^{(p-1)}\right]^{\mathrm{T}}\,\prod_{k=1}^{N-p} {t}^{N-(p-1),k}
R^{(p-1)}\,,
\label{e37}
\end{eqnarray}
which we will evaluate in an iterative manner as we did in
Sec. \ref{first2}. We define the  vector
$J^{N-p-1,N-p}={t}^{N-(p-1),N-p}R^{(p-1)}$ for $1<k<N-p$, and the
successive integrals are iteratively given by
\begin{eqnarray}
J^{N-p-1,k}=\int_{a_{N-p,k}}^{b_{N-p,k}}
dW_{N-p,k}\,{t}^{N-(p-1),k}\,J^{N-p-1,k+1}\,.
\label{it2}
\end{eqnarray}
The integral sign in Eq. (\ref{it2}) is interpreted in the sense that
we integrate each component of the vector integrand.  The final
integral (over $W_{N-p,1}$) yields, just as we saw in
Sec. \ref{first2},
$J_{N-p-1}=\left[L^{(p-1)}\right]^{\mathrm{T}}\,J^{N-p-1,1}$.

Step (ii): We then show that the $(i_1,i_2,\ldots,i_{p-1})$-th
component of $J^{N-p-1,k}$ is given by
\begin{eqnarray}
J^{N-p-1,k}_{i_1,\ldots,i_{p-1}}=\sum_{i_p=1}^{p+i_{p-1}}
\gamma^{N-p,k-1}_{i_1,\ldots,i_p}\,
(b_{N-p,k-1}-W_{N-p,k-1})^{p+j_{p-1}-j_p}.
\label{recform}
\end{eqnarray}
The vectors $\gamma^{N-p,k-1}$ and $\gamma^{N-p,k}$ with elements
$\gamma^{N-p,k-1}_{i_1,\ldots,i_p}$ and
$\gamma^{N-p,k}_{i_1,\ldots,i_p}$ respectively are related to each
other via the matrix relation
$\gamma^{N-p,k-1}={t}^{N-p,k}\,\gamma^{N-p,k}$.

Starting with $J^{N-p-1,N-p}$, we have
\begin{eqnarray}
J^{N-p-1,N-p}_{j1,\ldots,j_{p-1}}&=&\sum_{k_1,k_2,\ldots,
k_{p-1}}\beta_{j_1k_1,j_2k_2,\ldots,j_{p-1}
k_{p-1}}W^{p+j_{p-1}-k_{p-1}}_{N-p,N-p}\,R^{p-1}_{k_1,k_2,\ldots,
k_{p-1}}\nonumber\\ &=&\sum_{k_1,k_2,\ldots,
k_{p-1}}\beta_{j_1k_1,j_2k_2,\ldots,j_{p-1}
k_{p-1}}W^{p+j_{p-1}-1}_{N-p,N-p}\delta_{k_{p-1},1}R^{(p-1)}_{k_1,k_2,\ldots,k_{p-2}}\nonumber\\
&=&\sum_{j_p=1}^{p+j_{p-1}}\gamma^{N-p,N-p}_{j_1,j_2,\ldots,j_p}\,
(b_{N-p,N-p-1}-W_{N-p,N-p-1})^{p+j_{p-1}-j_p}\,,
\label{init}
\end{eqnarray}
where we have used Eq. (\ref{LR}) between the first and the second
lines of Eq. (\ref{init}), and the fact that
$W_{N-p,N-p}=b_{N-p,N-p-1}-W_{N-p,N-p-1}$ between the second and the
third lines. In other words, $J^{N-p-1,N-p}$ indeed has the postulated
form (\ref{recform} with
$\gamma^{N-p,N-p}_{j_1,\ldots,j_p}=\delta_{j_p,1}
\sum_{k_1,\ldots,k_{p-1}}\beta_{j_1k_1,\ldots,j_{p-1}
k_{p-1}}R^{(p-1)}_{k1,\ldots,k_{p-2}}$, where we choose $j_p$ to vary
between $1$ and $p+j_{p-1}$, as shown in Eq. (\ref{e36}).

Assuming the form (\ref{recform}) for $J^{N-p-1,k}$, we now calculate
$J^{N-p-1,k-1}$ using Eq. (\ref{it2}). The $(i_1,\ldots,i_{p-1})$-th
element of the corresponding integrand is
\begin{eqnarray}
\hspace{-1cm}\sum_{j_1,\ldots,j_p}\beta_{i_1j_1,\ldots,i_{p-1}
j_{p-1}}W^{p+i_{p-1}-j_{p-1}}_{N-p,k-1}
\gamma^{N-p,k-1}_{j_1,\ldots,j_p}
(b_{N-p,k-1}-W_{N-p,k-1})^{p+j_{p-1}-j_p}\,,
\label{intform}
\end{eqnarray}
so that after  using Eq. (\ref{Idev}), we obtain
\begin{eqnarray}
\hspace{-2cm}
  J^{N-p-1,k-1}_{i_1,\ldots,i_{p-1}}=\!\!\!\sum_{j_1,\ldots,j_{p}}\!\!\beta_{i_1j_1,\ldots,i_{p-1}
  j_{p-1}}
  \gamma^{N-p,k-1}_{j_1,\ldots,j_p}\,I_{p+i_{p-1}-j_{p-1},p+j_{p-1}-j_p}(a_{N-p,k-1},b_{N-p,k-1}).
\label{Jpp}
\end{eqnarray}
Using Eqs. (\ref{abprop1}) and (\ref{Idev}), we now expand
$I_{p+i_{p-1}-j_{p-1},p+j_{p-1}-j_p}(a_{N-p,k-1},b_{N-p,k-1})$ as
\begin{eqnarray}
\hspace{-1.5cm}I_{p+i_{p-1}-j_{p-1},p+j_{p-1}-j_p}(a_{N-p,k-1},b_{N-p,k-1})\nonumber\\
\hspace{0.5cm}=\sum_{l=0}^{p+i_{p-1}-j_{p-1}}
\alpha_{p+i_{p-1}-j_{p-1},p+j_{p-1}-j_p}^{p+i_{p-1}-j_{p-1}-l}\,
a_{N-p,k-1}^{p+i_{p-1}-j_{p-1}-l}\,
W_{N-p-1,k-1}^{p+1+j_{p-1}+l-j_p}\nonumber\\
\hspace{0.5cm}=\sum_{i_p=1}^{p+i_{p-1}}\Theta(i_p-j_{p-1})
\alpha_{p+i_{p-1}-j_{p-1},p+j_{p-1}-j_p}^{p+i_{p-1}-i_p}\,
a_{N-p,k-1}^{p+i_{p-1}-i_p}\, W_{N-p-1,k-1}^{p+1+i_p-j_p}\,,
\label{e43}
\end{eqnarray}
where in the last line of Eq. (\ref{e43}), $i_p=j_{p-1}
+l$. Thereafter, we rewrite Eq. (\ref{Jpp}) using Eq. (\ref{e43}) as
\begin{eqnarray}
\hspace{-2cm}J^{N-p-1,k-1}_{i_1,\ldots,i_{p-1}}=
\sum_{j_1,\ldots,j_{p}}\beta_{i_1j_1,\ldots,i_{p-1} j_{p-1}}
\gamma^{N-p,k-1}_{j_1,\ldots,j_p}\sum_{i_p=1}^{p+i_{p-1}}\Theta(i_p-j_{p-1})\times\nonumber\\
\hspace{2cm}\times\alpha_{p+i_{p-1}-j_{p-1},p+j_{p-1}-j_p}^{p+i_{p-1}-i_p}
a_{N-p,k-1}^{p+i_{p-1}-i_p} W_{N-p-1,k-1}^{p+1+i_p-j_p}\nonumber\\
=\!\!\!\sum_{i_p=1}^{p+i_{p-1}}\!\!\!\gamma^{N-p,k-2}_{i_1,\ldots,i_p}
a_{N-p,k-1}^{p+i_{p-1}-i_p}=\!\!\!\!\sum_{i_p}^{p+i_{p-1}}\!\!\!\!\gamma^{N-p,k-2}_{i_1,\ldots,i_p}
(b_{N-p,k-2}-W_{N-p,k-2})^{p+j_{p-1}-j_p}
\label{Ipp-1}
\end{eqnarray}
to recover
\begin{eqnarray}
\gamma^{N-p,k-2}_{i_1,\ldots,i_p}=\sum_{j_1,\ldots,j_{p}}
t^{N-p,k-1}_{i_1j_1,\ldots,i_{p}j_{p}}\gamma^{N-p,k-1}_{j_1,\ldots,j_p}
\label{e45}
\end{eqnarray}
and
\begin{eqnarray}
{
t}^{N-p,k-1}_{i_1j_1,\ldots,i_{p}j_{p}}&=&\beta_{i_1j_1,\ldots,i_{p-1}j_{p-1}}\alpha_{p+i_{p-1}-j_{p-1},p+j_{p-1}-j_p}^{p+i_{p-1}-i_p}\,\Theta(i_p-j_{p-1})\,W_{N-p-1,k-1}^{p+1+i_p-j_p}\nonumber\\
&=&\alpha_{1,2-j_1}^{2-i_1}\prod_{l=2}^{p}
\alpha_{l+i_{l-1}-j_{l-1},l+j_{l-1}-j_{l}}^{l+i_{l-1}-i_{l}}\Theta(i_l-j_{l-1})\,W^{p+1+i_{p}-j_{p}}_{N-p-1,k-1}\,.
\label{tform2}
\end{eqnarray}
Notice that ${t}^{N-p,k-1}$ indeed has the form postulated by
Eq. (\ref{tform}).

This procedure outlined in Eqs. (\ref{intform}-\ref{tform2}) allows us
to evaluate all the integrals in Eq. (\ref{e37}) up to $k=1$. Each
iteration in Eq. (\ref{e37}) introduces an extra multiplicative factor
of the matrix ${t}^{N-p,k}$, whose form is given in Eq. (\ref{tform2}).

For the last integration (over $W_{N-p,1}$) in Eq. (\ref{e37}),
$a_{N-p,1}=0$ and $b_{N-p,1}=1$, so that
$J^{N-p-1,1}_{i_1,\ldots,i_p-1}$ is simply equal to
$\gamma^{N-p,1}_{i_1,i_2,\ldots,i_{p-1},p+i_{p-1}}$; i.e.,
\begin{eqnarray}
J_{N-p-1}&=&\sum_{i_1,i_2,\ldots i_{p-1}}L_{i_1,i_2,\ldots,
i_{p-1}}^{(p-1)}\gamma^{N-p,1}_{i_1,i_2,\ldots,
i_{p-1},p+i_{p-1}}\nonumber\\&=&\sum_{i_1,i_2,\ldots,
i_{p}}L_{i_1,i_2,\ldots,
i_{p-1}}^{p-1}\delta_{i_p,p+i_{p-1}}\gamma^{N-p,1}_{i_1,i_2,\ldots,
i_{p-1},i_p}\nonumber\\&=&\sum_{i_1,i_2,\ldots,
i_{p}}L_{i_1,i_2,\ldots,
i_{p-1}}^{(p-1)}\delta_{i_p,p+i_{p-1}}\left[\prod_{k=1}^{N-p-1}{t}^{N-p,k}\gamma^{N-p,N-p}\right]_{i_1,i_2,\ldots,
i_{p-1},i_p}\nonumber\\&=&\left[L^{(p)}\right]^{\mathrm{T}}
\prod_{k=1}^{N-p-1}{t}^{N-p,k} R^{(p)}\,,
\label{e47}
\end{eqnarray}
where
\begin{eqnarray}
L^{(p)}_{i_1,\ldots,i_{p}}=L_{i_1,\ldots,i_{p-1}}^{(p-1)}\delta_{i_p,p+i_{p-1}}=\prod_{l=1}^{p}\delta_{i_l,\frac{l(l+1)}{2}+1}
\label{e48}
\end{eqnarray}
and
\begin{eqnarray}
R^{(p)}_{j_1,j_2,\ldots,j_{p}}=\gamma^{N-p,N-p}_{j_1,j_2,\ldots,j_{p}}=\delta_{j_p,1}
\sum_{k_1,k_2,\ldots,k_{p-1}}\beta_{j_1k_1,j_2k_2,\ldots,j_{p-1}
k_{p-1}}R^{(p-1)}_{k_1,\ldots,k_{p-1}}\,;
\label{e49}
\end{eqnarray}
i.e., Eqs. (\ref{e48}-\ref{e49}) are exactly of the form postulated in
Eq. (\ref{LR}).

\subsubsection{The normalization constant ${\cal N}$:\label{sec4.2.3}}

The above induction procedure allows us to evaluate $J_{N-p}$ all the
way to $p=N-1$, and the expression for the normalization constant is
then given by
\begin{eqnarray}
{\cal
N}=J_{1}=\left[L^{(N-2)}\right]^{\mathrm{T}}\,{t}^{2,1}\,R^{(N-2)}\,.
\label{norm}
\end{eqnarray}
For the full expression of ${\cal N}$ however, we also need to use
Eqs. (\ref{alpha}), (\ref{tform}) and (\ref{beta}) to obtain
\begin{eqnarray}
{ t}^{2,1}_{i_1j_1,i_2j_2,\ldots,i_{N-2}
j_{N-2}}=\beta_{i_1j_1,i_2j_2,\ldots,i_{N-2} j_{N-2}}\qquad\mbox{since
$W_{1,1}=W_{\mathrm{ext}}=1$}\,,
\label{e51}
\end{eqnarray}
\begin{eqnarray}
\hspace{-2cm}\beta_{i_1j_1,\ldots,i_{p}j_p}\!=\!\frac{(2\!-\!j_1)!}{(2\!-\!i_1)!}\frac{(2\!+\!j_1\!-\!j_2)!}{(2\!+\!i_1\!-\!i_2)!}\ldots\frac{(p\!+\!j_{p-1}\!-\!j_p)!}{(p\!+\!i_{p-1}\!-\!i_p)!}\,\frac{1}{(p\!+\!1\!+\!i_p\!-\!j_p)!}\,\prod_{l=2}^{p-1}\Theta(i_l\!-\!j_{l\!-\!1}),
\label{betaform}
\end{eqnarray}
and further Eqs. (\ref{alpha}) and (\ref{LR}) to derive
\begin{eqnarray}
R^{(p)}_{i_1,\ldots,i_{p}}=
\frac{\Lambda^{(p)}_{i_1\ldots,i_{p}}}{(2\!-\!i_1)!(2\!+\!i_1\!-\!i_2)!\ldots(p\!-\!1\!+\!i_{p-2}\!-\!i_{p-1})!(p\!-\!1\!+\!i_{p-1})!}\,.
\label{Rform}
\end{eqnarray}
In Eq. (\ref{Rform}), $\Lambda^{(p)}_{j_1\ldots j_{p}}$ are integers
given by the following recursive formula:
\begin{eqnarray}
\Lambda^{(2)}_{j_1j_2}&=&\delta_{j_2,1}\nonumber\\
\Lambda^{(p+1)}_{j_1,\ldots,j_{p+1}}&=&\delta_{j_{p+1},1}\sum_{k_1,\ldots,k_p}\Lambda^{(p)}_{k_1,\ldots,k_{p}}\prod_{l=1}^{p}\Theta(j_{l+1}-k_l)\,.
\label{lambda_form}
\end{eqnarray}
Finally, using Eqs. (\ref{norm}-\ref{lambda_form}) and (\ref{e48}), we
have
\begin{eqnarray}
{\cal
N}=J_{1}=\frac{1}{\left[\frac{(N-1)N}{2}\right]!}\,\,\sum_{j_1,\ldots,j_{N-2}}
\Lambda^{(N-2)}_{j_1,\ldots,j_{N-2}}\,.
\label{e55}
\end{eqnarray}

\subsection{Calculation of the Expectation Values $\langle
  W_{N-q,r}\rangle$ \label{expect}}

\subsubsection{Modifications of fundamental relations:\label{mean_sect}}

\begin{figure*}
\begin{center}
\includegraphics[width=0.45\linewidth]{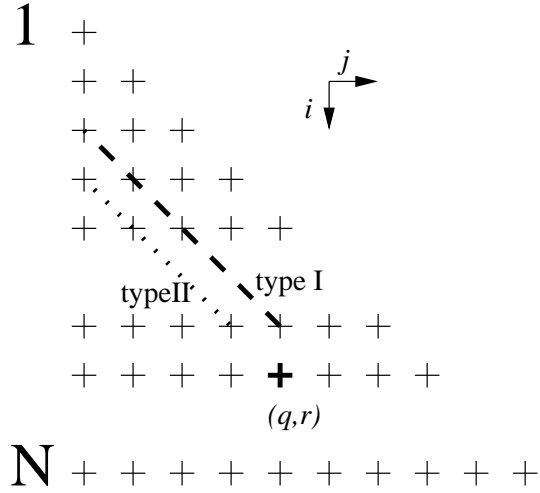}
\caption{Propagation of defects in the recurrence relations for the
  calculation of $\langle W_{N-q,r}\rangle$. \label{mesh2}}
\end{center}
\end{figure*}

In this section, we obtain the general form of $\langle W_{N-q,r}
\rangle$, the mean value of $W_{N-q,r}$ at $q$ layers from the bottom
and $r$ grains from the left boundary [see Fig. \ref{mesh2} for
illustration], for $0\leq q\leq N-1$ and $1\leq r\leq N-q-1$, in four
steps. From what we learnt in Sec. \ref{norm_sect}, it is clear that
in order to calculate $\langle W_{N-q,r}\rangle$, we need to integrate
$W_{N-q,r}J_{N-q}$ from the $(N-q)$-th layer to all the way to the
top. Just like we saw in Sec. \ref{norm_sect}, these integrations are
equivalent to matrix multiplications layer by layer in an iterative
manner, but the fundamental relations (\ref{tform}-\ref{usual}) do get
slightly modified.

Step (i): We start with the calculation of the integral of
$W_{N-q,r}J_{N-q}$ on the $(N-q)$-th layer
\begin{eqnarray}
\hspace{-1cm}\bar{J}_{N-q-1}(q,r)=\int\left[\prod_{l'=1}^{N-q-1}dW_{N-q,l'}\right]W_{N-q,r}\left[L^{(q-1)}\right]^{\mathrm{T}}\left[\prod_{l=1}^{N-q}{t}^{N-(q-1),l}\right]R^{(q-1)}\!,
\label{meanfirst}
\end{eqnarray}
wherein once again we carry out the integrations iteratively from
$l'=N-q-1$ down to $l'=1$. Of them, notice that the integrations for
$l'=N-q-1$ down to $l'=r+1$ proceed exactly as in
Eqs. (\ref{recform}-\ref{tform2}), yielding
$\bar{J}^{N-q-1,r+1}(q,r)\equiv J^{N-q-1,r+1}$, i.e.,
\begin{eqnarray}
\hspace{-2cm}\bar{J}_{N-q-1}(q,r)&=&\int\left[\prod_{l'=1}^{r}dW_{N-q,l'}\right]W_{N-q,r}\left[L^{(q-1)}\right]^{\mathrm{T}}\left[\prod_{l=1}^{r}{t}^{N-(q-1),l}\right]J^{N-q-1,r+1}(q,r)\nonumber\\
&\equiv&\int\left[\prod_{l'=1}^{r-1}dW_{N-q,l'}\right]\left[L^{(q-1)}\right]^{\mathrm{T}}\left[\prod_{l=1}^{r-1}{t}^{N-(q-1),l}\right]\bar{J}^{N-q-1,r}(q,r)\,.
\label{it_mod}
\end{eqnarray}

Step (ii): For $l'=r$,  the recursive integration procedure differs
from those in Eqs. (\ref{recform}-\ref{tform2}) due to the presence of
an extra factor of $W_{N-q,r}$ in the integrand, and thus the
integration over $W_{N-q,r}$ in Eq. (\ref{meanfirst}) requires a
slightly different treatment.We first write $\bar{J}^{N-q-1,r}(q,r)$
explicitly:
\begin{eqnarray}
\hspace{-2cm}\bar{J}^{N-q-1,r}_{i_1,\ldots,i_{q-1}}(q,r)=\int_{a_{N-q,r}}^{b_{N-q,r}}dW_{N-q,r}
W_{N-q,r}\sum_{j_1,\ldots,j_q}\beta_{i_1j_1,\ldots,i_{q-1}j_{q-1}}W^{q+i_{q-1}-j_{q-1}}_{N-q,r}\times\nonumber\\
\hspace{1.5cm}\times\,\gamma^{N-q,r+1}_{j_1,\ldots,j_q}
(b_{N-q,r}-W_{N-q,r})^{q+j_{q-1}-j_q}\nonumber\\
\hspace{5mm}=\!\!\!\sum_{j_1,\ldots,j_q}\!\!\beta_{i_1j_1,\ldots,i_{q-1}j_{q-1}}\gamma^{N-q,r+1}_{j_1,\ldots,j_q}I_{q+1+i_{q-1}-j_{q-1},q+j_{q-1}-j_q}(a_{N-q,r},b_{N-q,r})\,.
\label{meanint1}
\end{eqnarray}
We then use
\begin{eqnarray}
\hspace{-1cm}I_{q+1+i_{q-1}-j_{q-1},q+j_{q-1}-j_q}(a_{N-q,r},b_{N-q,r})\nonumber\\
\hspace{0cm}=\sum_{l=0}^{q+1+i_{q-1}-j_{q-1}}
\alpha_{q+1+i_{q-1}-j_{q-1},q+j_{q-1}-j_q}^{q+1+i_{q-1}-j_{q-1}-l}
\,a_{N-q,r}^{q+1+i_{q-1}-j_{q-1}-l}\,W_{N-q-1,r}^{q+1+j_{q-1}+l-j_q}\nonumber\\
\hspace{0cm}=\sum_{i_q=1}^{q+1+i_{q-1}}\Theta(i_q-j_{q-1})
\,\alpha_{q+1+i_{q-1}-j_{q-1},q+j_{q-1}-j_q}^{q+1+i_{q-1}-i_q}
\,a_{N-q,r}^{q+1+i_{q-1}-i_q}\,W_{N-q-1,r}^{q+1+i_q-j_q}
\label{e59}
\end{eqnarray}
(where once again we introduce $i_q=j_{q-1}+l$), to get
\begin{eqnarray}
\bar{J}^{N-q-1,r}_{i_1,\ldots,i_{q-1}}(q,r)=\sum_{i_q=1}^{q+1+i_{q-1}}
\bar{\gamma}^{N-q,r}_{i_1,\ldots,i_q}\,(b_{N-q,r-1}-W_{N-q,r-1}
)^{q+1+i_{q-1}-i_q}
\label{mean_first_res}
\end{eqnarray}
with $\bar{\gamma}^{N-q,r}={\bar{t}(q,r)}^{N-q,r}\,\gamma^{N-q,r+1}$,
where
\begin{eqnarray}
\hspace{-1cm}\bar{t}^{N-q,r}_{i_1j_1,\ldots,i_{q}j_{q}}(q,r)&=&
\beta_{i_1j_1,\ldots,i_{q-1}j_{q-1}}\Theta(i_q-j_{q-1})\,\alpha_{q+1+i_{q-1}-j_{q-1},q+j_{q-1}-j_q}^{q+1+i_{q-1}-i_q}
\,W_{N-q-1,r}^{q+1+i_q-j_q}\nonumber\\
&=&\bar{\beta}_{i_1j_1,\ldots,i_{q}j_{q}}^{q,r}\,W_{N-q-1,r}^{q+1+i_q-j_q}\,.
\label{tbar}
\end{eqnarray}
It is worthwhile to note that all indices except $i_q$ in
Eqs. (\ref{e59}-\ref{tbar}) run between the bounds defined in
(\ref{ind_usual}), while for $i_q$, we have $1\leq i_q\leq q+1+
i_{q-1}$. In other words, the fundamental relation (\ref{tform}) gets
modified, as $\bar{t}^{N-q,r}$ has one more row than
${t}^{N-q,r}$. Such a modification can be thought of as a ``defect''
in the recursive integration procedure
(\ref{recform}-\ref{tform2}). More precisely, we call Eq. (\ref{tbar})
a ``defect of type I at $(q,r)$'' and it is characterized by the
following relation between $\beta$ and $\bar\beta^{q,r}$ [this
relation is obtained via the usage of Eq. (\ref{alpha})]:
\begin{eqnarray}
\bar{\beta}_{i_1j_1,\ldots,i_{q}j_{q}}^{q,r}=\beta_{i_1j_1,\ldots,i_{q}j_{q}}\frac{q+1+i_{q-1}-j_{q-1}}{q+1+i_{q-1}-i_{q}}\,.
\label{defec1}
\end{eqnarray}

Step (iii): Next we show that one needs yet another recursive form
when $l'=r-1$ in Eq. (\ref{meanfirst}), but thereafter from $l'=r-2$
down to $l'=1$, the recursive integration scheme of
Eq. (\ref{meanfirst}) returns to its old form
(\ref{recform}-\ref{tform2}).

The recursive form for $l'=r-1$ is easily obtained by taking the
matrix product of  ${t}^{N-(q-1),r-1}$ and $\bar{J}^{N-q-1,r}(q,r)$
and then integrating it w.r.t. $W_{N-q,r-1}$, yielding
\begin{eqnarray}
\hspace{-2.3cm}\bar{J}^{N-q-1,r-1}_{i_1\ldots
i_{q-1}}(q,r)=\!\!\!\sum_{j_1,\ldots,j_q}\!\!\!\beta_{i_1j_1,\ldots,i_{q-1}j_{q-1}}\,\bar{\gamma}^{N-q,r}_{j_1,\ldots,j_q}\,I_{q+i_{q-1}-j_{q-1},q+1+j_{q-1}-j_q}(a_{N-q,r-1},b_{N-q,r-1})
\nonumber \\
\hspace{0.6cm}=\sum_{i_q=1}^{q+i_{q-1}}\bar{\gamma}^{N-q,r-1}_{i_1,\ldots,i_q}\,(b_{N-q,r-2}-W_{N-q,r-2})^{q+i_{q-1}-i_q}\,,
\label{meansec}
\end{eqnarray}
where the last line of Eq. (\ref{meansec}) has been obtained by using
Eq. (\ref{Idev}). Thereafter, using Eq. (\ref{alpha}), we get
$\bar{\gamma}^{N-q,r-1}={{\bar{t}}}^{N-q,r-1}(q,r)\bar{\gamma}^{N-q,r}$,
with
\begin{eqnarray}
\hspace{-2cm}\bar{t}^{N-q,r-1}_{i_1j_1,\ldots,i_{q}j_{q}}(q,r)=\beta_{i_1j_1,\ldots,i_{q-1}j_{q-1}}\,\Theta(i_q-j_{q-1})\,\alpha_{q+i_{q-1}-j_{q-1},q+1+j_{q-1}-j_q}^{q+i_{q-1}-i_q}
\,W_{N-q-1,r-1}^{q+2+i_q-j_q}\,.
\label{tbarbar}
\end{eqnarray}
In Eq. (\ref{tbarbar}), all indices except $j_q$ satisfy
Eq. (\ref{ind_usual}), and for $j_q$, we have $1\leq j_q\leq
q+1+j_{q-1}$; i.e., the fundamental relation is once again modified:
$\bar{t}^{N-q,r-1}(q,r)$ has one more column than ${t}^{N-q,r-1}$. We
call this modification a ``defect of type II at $(q,r-1)$'', which is
characterized by the following relation between $\beta$ and
$\bar\beta^{q,r-1}$ [once again, obtained via the usage of
Eq. (\ref{alpha})]:
\begin{eqnarray}
\bar{\beta}_{i_1j_1,i_2j_2,\ldots
i_{q}j_{q}}^{q,r-1}=\beta_{i_1j_1,i_2j_2,\ldots
i_{q}j_{q}}\frac{q+1+j_{q-1}-j_{q}}{q+2+i_{q}-j_{q}}\,.
\label{defec2}
\end{eqnarray}
As for the integrations in Eq. (\ref{meanfirst}) for $l'=r-2$ down to
$l'=1$, notice that the final expression (\ref{meansec}) is the same
as the earlier expression (\ref{init}), although the matrices relating
the ${\gamma}$'s have been modified. Thus, the rest of the
integrations in Eq. (\ref{meanfirst}) for the $(N-q)$-th layer yield
the recurrence relation (\ref{recform}), and the fundamental relations
for the matrices remain the same as in Eq. (\ref{tform}). The final
result then simply becomes
\begin{eqnarray}
\bar{J}_{N-q-1}=\!\left[L^{(q)}\right]^{\mathrm T}
\prod_{l'=1}^{r-2}{t}^{N-q,l'}\,\bar{t}^{N-q,r-1}(q,r)\,\bar{t}^{N-q,r}(q,r)\prod_{l=r+1}^{N-q-1}{t}^{N-q,l}
R^{(q)},
\label{meanfirstlay_res}
\end{eqnarray}
i.e., in the $(N-q)$-th layer, only two of the matrices get modified
w.r.t. Eq. (\ref{tform}).

Step (iv): In the fourth step, we integrate $\bar{J}_{N-q-1}$ for the
remaining $N-q-2$ layers to the top of the pile one by one. This
integration procedure is the same as what has been detailed in steps
(i)-(iii), so here we only provide a short description of it.

When we integrate $\bar{J}_{N-q-1}$ in the $(N-q-1)$-th layer, it is
easily seen [following the calculations in steps (i)-(iii) above] that
the fundamental relations for the matrices remain the same as in
Eq. (\ref{tform}) for the integrations over $W_{N-q-1,N-q-2}$ down to
$W_{N-q-1,r}$. Two defects, one of type I and another of type II
appear respectively at locations $(N-q-1,r-1)$ and at $(N-q-1,r-2)$, but
the fundamental relations (\ref{tform}) are recovered for the
integrations over $W_{N-q-1,r-3}$ down to $W_{N-q-1,1}$. In other
words, after the integrations of $\bar{J}_{N-q-1}$ over all the
$W_{N-q-1,j}$'s for $j=(N-q-2),\ldots,1$, the locations of the defects
move one grain each towards the left. In fact, this trend of the
leftward shift of the defects by one grain each time we integrate over
all the $W_{i,j}$'s in the successive layer upwards continues to hold
until the locations of the defects reach (and terminate on) the left
boundary [see Fig. \ref{mesh2}].

In summary, effectively, the difference between the calculations of
${\cal N}$ and that of $\langle W_{N-q,r}\rangle$ lies in the fact
that for the latter calculation, the fundamental relations between the
matrices are modified on the grains located at $(N-q-k,r-k)$, $0\leq k
\leq r-1$ and $(N-q-k,r-1-k)$, $0\leq k \leq r-2$. The final result is
of the form
\begin{eqnarray}
\langle W_{N-q,r} \rangle=\frac{1}{\cal
N}\left[\bar{L}^{(N-2)}\right]^{\mathrm T}\,\bar{t}^{
2,1}(q,r)\,\bar{R}^{(N-2)}(q,r)\,,
\label{meanform}
\end{eqnarray}
but the explicit form of $\langle W_{N-q,r} \rangle$ depends on how
the modified fundamental relations affect $\bar{L}^{(N-2)}$,
$\bar{t}^{2,1}(q,r)$, and $\bar{R}^{(N-2)}(q,r)$, and hence on the
values of $r$.

\subsubsection{Calculation of $\langle W_{N-q,r}\rangle$ on the
boundary, i.e., $r=1$:\label{sec4.3.2}}

For $r=1$, there is only one relevant defect, and it is of type I. It
affects only $\bar{L}^{(N-2)}$ and $\bar{t}^{2,1}(q,1)$, while
$\bar{R}^{(N-2)}(q,1)$ remains the same as ${R}^{(N-2)}$. Let us first
consider the case $q>1$, for which we have
\begin{eqnarray}
\bar{t}^{2,1}(q,1)=\bar{\beta}_{i_1j_1,\ldots,i_{N-2}
j_{N-2}}^{(q,1)}\,.
\label{e68}
\end{eqnarray}
Herein, the fundamental relation for
$\bar{\beta}_{i_1j_1,\ldots,i_{N-2}j_{N-2}}(q,1)$ gets modified only
at the $q$-th layer to the form (\ref{defec1}) to yield
\begin{eqnarray}
\bar{\beta}_{i_1j_1,\ldots,i_{N-2}j_{N-2}}^{q,1}=\beta_{i_1j_1,\ldots,i_{N-2}j_{N-2}}\frac{q+1+i_{q-1}-j_{q-1}}{q+1+i_{q-1}-i_{q}}\,.
\label{defec1_boundary}
\end{eqnarray}
In addition, we should also keep in mind that the dimensions of the
matrix $\bar{t}^{2,1}(q,1)$ are not the same as those of ${t}^{2,1}$,
since the index $i_q$ for $\bar{t}^{2,1}(q,1)$ varies between $1$ and
$q+1+i_{q-1}$ as opposed to varying between $1$ and $q+i_{q-1}$ for
${t}^{2,1}$. This implies that the maximum value attained by $i_l$ for
$\bar{t}^{2,1}(q,1)$ for $l\geq q$ is increased by $1$ due to the
presence of the defect of type I at location $(N-q,r)$, and
consequently
\begin{eqnarray}
\bar{L}_{i_1,\ldots,i_{N-2}}^{(N-2)}&=&\prod_{l'=1}^{q-1}\delta_{i_{l'},\frac{l'(l'+1)}{2}+1}\prod_{l=q}^{N-2}\delta_{i_l,\frac{l(l+1)}{2}+2}
\label{e70}
\end{eqnarray}
When Eqs. (\ref{e68}-\ref{e70}) are put together along with
Eqs. (\ref{usual}) and (\ref{alpha}) and (\ref{lambda_form}) in
Eq. (\ref{meanform}), we obtain
\begin{eqnarray}
\langle W_{N-q,1}\rangle=\frac{1}{\cal
N}\sum_{j_1,\ldots,j_{N-2}}\frac{\left[\frac{q(q+1)}{2}+2-j_{q-1}\right]\Lambda^{(N-2)}_{j_1,\ldots,j_{N-2}}}{\left[\frac{N(N-1)}{2}+1\right]!}\,.
\label{boundary_final}
\end{eqnarray}

The two cases $q=0$ and $q=1$, however, have to be considered
separately. For $q=1$, we find that  Eq. (\ref{boundary_final}) can be
generalized by using $j_0=1$, yielding
\begin{eqnarray}
\langle W_{N-1,1} \rangle=\frac{4}{(N-1)N+1}
\label{e73a}
\end{eqnarray}
while for $q=1$, it can be seen that $\langle W_{N,1} \rangle=\langle
W_{N-1,1} \rangle/2$.

\subsubsection{In the bulk, i.e., $r>1$:\label{sec4.3.3}} For $r>1$, each quantity
in Eq. (\ref{meanform}) is affected due to the development of the
defects. Of these quantities, $\bar{t}^{\,2,1}(q,r)$ is affected by a
defect of type I terminating on the boundary at location $(q+r-1,1)$
and a defect of type II at location $(q+r-2,1)$. Using
Eqs. (\ref{defec1}) and (\ref{defec2}), we then get
\begin{eqnarray}
\bar{\beta}_{i_1j_1,\ldots,i_{N-2}j_{N-2}}^{q,r}
=\beta_{i_1j_1,\ldots,i_{N-2}j_{N-2}}\frac{q+r-1+j_{q+r-3}-j_{q+r-2}}{q+r+i_{q+r-2}-i_{q+r-1}}\,.
\label{e73}
\end{eqnarray}
We also obtain, just as before,
\begin{eqnarray}
\bar{L}_{i_1,\ldots,i_{(N-2)}}^{(N-2)}&=&\prod_{l'=1}^{q+r-2}\delta_{i_{l'},\frac{l'(l'+1)}{2}+1}\prod_{l=q+r-1}^{N-2}\delta_{i_l,\frac{l(l+1)}{2}+2}\,.
\label{e74}
\end{eqnarray}

However, we still need to express $\bar{R}^{(N-2)}(q,r)$. Recall from
Eq.(\ref{LR}) that  ${R}^{(N-2)}$ depends, through the recursion
formula, on ${t}^{N-p, N-p-1}$ for $1\leq p\leq N-1$. Since for
the calculation of $\langle W_{N-q,r} \rangle$, ${t}^{N-p, N-p-1}$
for $N-r-1 \leq p \leq N-3$ are modified due to the defects,
$\bar{R}^{(N-2)}(q,r)$ must also differ from ${R}^{(N-2)}$.
  
More precisely, in the recursive formalism described above,
$\bar{R}^{(p)}(q,r)$ does not differ from ${R}^{(p)}(q,r)$ for
$p \leq N-r-2$. The value $p=N-r-1$ corresponds to the layer number where a
vertical line drawn from $(q,r)$ in Fig. \ref{mesh2} intersects the
right edge of the triangle.

Thus, up to $p=N-r-2$, the usual recursion formula applies so that 
\begin{eqnarray}
\hspace{-2.5cm}R^{(N-r-1)}_{i_1, \ldots, i_{N-r-1}}\!\!=\!
\frac{\Lambda^{(N-r-1)}_{i_1,\ldots,i_{N-r-1}}}{(2\!-\!i_1)!(2\!+\!i_1\!-\!i_2)!...(N\!-\!r\!-\!2\!+\!i_{N-r-3}\!-\!i_{N-r-2})!(N\!-\!r\!-\!2\!+\!i_{N-r-2})!},
\end{eqnarray}
but then $\bar{\beta}^{r+1, r}$ is modified with respect to
${\beta}^{r+1, r}$ by a defect of type I at $(q,r)$, and we get
\begin{eqnarray}
\hspace{-2.2cm}\bar{R}^{(N-r-1)}_{i_1,\ldots, i_{N-r}}\!\!=\!
\frac{\frac{1}{q+1+i_{q-1}-i_q}\bar{\Lambda}^{(N-r)}_{i_1\ldots
i_{N-r}}(q,r)}{(2\!-\!i_1)!(2\!+\!i_1\!-\!i_2)!...(N\!-\!r\!-\!1\!+\!i_{N-r-2}\!-\!i_{N-r-1})!(N\!-\!r\!-\!1\!+\!i_{N-r-1})!},
\end{eqnarray}
where
\begin{eqnarray}
\hspace{-1cm}\bar{\Lambda}^{(N-r)}_{i_1,\ldots,i_{N-r}}(q,r)=\sum_{j_1,\ldots,j_{N-r-1}}(q+1+i_{q-1}-j_{q-1})\,\prod_{l}
\Theta(i_l-j_{l-1})\,\Lambda^{(N-r-1)}_{j_1\ldots j_{N-r-1}}.
\label{Rdefect}
\end{eqnarray}
Once again, it is important to remember  that the $q$-th index $i_q$
satisfies $1\leq i_q\leq q+1+i_{q-1}$, while all the others satisfy
the inequalities (\ref{e36}).

At the next step of the recursion formula for $R^{(N-2)}$,
$\bar{\beta}^{r, r-1}$ is modified with respect to ${\beta}^{r, r-1}$
by two defects: one of type I at $(q+1,r-1)$ and one of type II at
$(q,r-1)$ and we obtain
\begin{eqnarray}
\hspace{-2.3cm}\bar{R}^{(N-r+1)}_{i_1,\ldots, i_{N-r+1}}  & = &
\frac{\frac{1}{q+2+i_{q}-i_{q+1}}\bar{\Lambda}^{(N-r+1)}_{i_1\ldots
i_{N-r+1}}(q,r)}{(2-i_1)!(2+i_1-i_2)!\ldots
(N-r+i_{N-r-1}-i_{N-r})!(N-r+i_{N-r})!},
\end{eqnarray}
where
\begin{eqnarray}
\bar{\Lambda}^{(N-r+1)}_{i_1,\ldots,
i_{N-r+1}}(q,r)=\sum_{j_1,\ldots,j_{N-r}}\prod_{l}
\Theta(i_l-j_{l-1})\,\bar{\Lambda}^{(N-r)}_{j_1,\ldots,j_{N-r}}(q,r),
\label{80}
\end{eqnarray}
with $1\leq j_{q}\leq q+1+j_{q-1}$ and $1\leq i_{q+1}\leq q+2+i_{q}$.
Notice that Eq. (\ref{80}) is of the same form as
Eq. (\ref{lambda_form}) except for the bounds of $j_q$ and $i_{q+1}$.

The calculations for $\bar{R}^{(N-k)}(q,r)$, $2\leq k \leq r$ proceed
along the same lines as for $\bar{R}^{(N-r+1)}(q,r)$ and finally we
obtain
\begin{eqnarray}
\hspace{-2cm}\bar{R}^{(N-2)}_{i_1,\ldots, i_{N-3}}  & = &
\frac{\frac{1}{q+r-1+i_{q+r-3}-i_{q+r-2}}\bar{\Lambda}^{(N-2)}_{i_1,\ldots,i_{N-2}}(q,r)}{(2-i_1)!(2-i_1-i_2)!\ldots(N-4+i_{N-5}-i_{N-4})!(N-4+i_{N-4})!},
\end{eqnarray}
where $1\leq i_{q+r-2}\leq q+r-1+i_{q+r-3}$, and for all the other
indices, we have  $1\leq i_l\leq l+i_{l-1}$.

Putting everything together, as usual, most of the factorials that
appear in the expressions of $\bar{t}^{2,1}(q,r)$ and
$\bar{R}^{(N-2)}(q,r)$ cancel out and we are left with
\begin{eqnarray}
\langle W_{N-q,r}
\rangle=\frac{1}{\cal{N}}\sum_{j_1,\ldots,j_{N-2}}\frac{\bar{\Lambda}^{(N-2)}_{j_1,\ldots,j_{N-2}}(q,r)}{\left[\frac{N(N-1)}{2}+1\right]!}.
\label{mean_fin}
\end{eqnarray}

\subsection{Higher moments and correlations\label{sec4.4}}

From the methods we described in Secs. \ref{prelim}-\ref{expect}, it
is clear that  higher moments and correlations can in principle be
calculated by following the same procedure.  We will not provide too
many details below, instead we will demonstrate  that the higher
moments and correlations can easily be obtained  by keeping track of
more ``defects'' in the fundamental relations.

\subsubsection{The case of $\langle W^{s}_{N-q,r}\rangle$, $s>1$}
The modifications of the fundamental relations for the integration of
$W_{N-q,r}^s$ are obtained  by generalizing the calculations of
section \ref{mean_sect}.

To start with, the $(i_1,\ldots, i_{q-1})$-th element of the matrix
integration result $\displaystyle {\int_{a_{N-q,r}}^{b_{N-q,r}}}
dW_{N-q,r}W_{N-q,r}^{s}{t}^{N-(q-1),k}{\bar{J}}^{N-q-1,r+1}(q,r)$ reads
\begin{eqnarray}
\nonumber
\hspace{-1cm}  \sum_{j_1,\ldots,j_q}\beta_{i_1j_1,\ldots,i_{q-1}j_{q-1}}\gamma^{N-q,r+1}_{j_1,\ldots,j_q}I_{q+s+i_{q-1}-j_{q-1},q+j_{q-1}-j_q}(a_{N-q,r},b_{N-q,r})\\
\hspace{1.5cm}=\sum_{i_q=0}^{q+s+i_{q-1}}
\bar{\gamma}^{N-q,r,s}_{i_1,\ldots
i_q}(b_{N-q,r-1}-W_{N-q,r-1})^{q+s+i_{q-1}-i_q}
\label{sint1}
\end{eqnarray}
with
$\bar{\gamma}^{N-q,r,s}={\bar{{t}}(q,r,s)}^{N-q,r}\gamma^{N-q,r+1}$,
where
\begin{eqnarray}
\nonumber
\hspace{-2cm}\bar{t}^{N-q,r}_{i_1j_1,\ldots,i_{q}j_{q}}(q,r,s) & = &
\beta_{i_1j_1,\ldots,i_{q-1}j_{q-1}}\Theta(i_q-j_{q-1})\,\alpha_{q+s+i_{q-1}-j_{q-1},q+j_{q-1}-j_q}^{q+s+i_{q-1}-i_q}\,W_{N-q-1,r-1}^{q+1+i_q-j_q}\\
&  = &
\bar{\beta}_{i_1j_1,\ldots,i_{q}j_{q}}^{(q,r,s)}W_{N-q-1,r-1}^{q+1+i_q-j_q}.
\end{eqnarray}

Once again, all the indices vary between the bounds defined in
(\ref{ind_usual}), except for $i_q$, which satisfies $1\leq i_q \leq
q+s+i_{q-1}$. The resulting modification of the fundamental relations
can be called an $s$-th order defect of type I, and it is
characterized by the relation

\begin{eqnarray}
\bar{\beta}_{i_1j_1,\ldots,i_{q}j_{q}}^{q,r,s}=\beta_{i_1j_1,\ldots,i_{q}j_{q}}\frac{(q+s+i_{q-1}-j_{q-1})\ldots(q+1+i_{q-1}-j_{q-1})}{(q+s+i_{q-1}-i_{q})\ldots(q+1+i_{q-1}-i_{q})}.
\end{eqnarray}

After multiplying the r.h.s. of Eq.({\ref{sint1}}) with
${t}^{N-(q-1),r-1}$, the integral with respect to $W_{N-q,r-1}$ yields
\begin{eqnarray}
\nonumber
\hspace{-1cm}\sum_{j_1,\ldots,j_q}\beta_{i_1j_1,\ldots,i_{q-1}j_{q-1}}\,\bar{\gamma}^{N-q,r,s}_{j_1,\ldots,j_q}\,I_{q+i_{q-1}-j_{q-1},q+s+j_{q-1}-j_q}(a_{N-q,r-1},b_{N-q,r-1})\\
\hspace{2cm}=\sum_{i_q=0}^{q+i_{q-1}}
\bar{\gamma}^{N-q,r-1,s}_{i_1\ldots
i_q}(b_{N-q,r-2}-W_{N-q,r-2})^{q+i_{q-1}-i_q}.
\label{sint2}
\end{eqnarray}
Furthermore, using
$\bar{\gamma}^{N-q,r-1,s}={\bar{t}}^{N-q,r-1}(q,r,s)\,\bar{\gamma}^{N-q,r,s}$,
we obtain
\begin{eqnarray}
\hspace{-2cm}\bar{{t}}^{N-q,r-1,s}_{i_1j_1,\ldots,i_{q}j_{q}}(q,r)=\beta_{i_1j_1,\ldots,i_{q-1}j_{q-1}}\Theta(i_q-j_{q-1})\,\alpha_{q+i_{q-1}-j_{q-1},q+s+j_{q-1}-j_q}^{q+i_{q-1}-i_q}
W_{N-q-1,r-1}^{q+s+1+i_q-j_q},
\end{eqnarray}
where  $1\leq j_q \leq s+q+j_{q-1}$, while all the other indices
satisfy Eq.(\ref{ind_usual}). Thus, the $s$-th order defect of type II
is characterized by
\begin{eqnarray}
\bar{\beta}_{i_1j_1,\ldots,i_{q}j_{q}}^{(q,r-1,s)}=\beta_{i_1j_1,\ldots,i_{q}j_{q}}\frac{(q+s+j_{q-1}-j_{q})\ldots(q+1+j_{q-1}-j_{q})}{(q+s+1+i_{q}-j_{q})\ldots(q+2+i_{q}-j_{q})}.
\label{e88}
\end{eqnarray}
As in the case for $s=1$, all the other integrals on the $(N-q)$-th
layer have the usual form (\ref{recform}), so that there are only two
defects per layer to take care of. Moreover, these defects propagate
exactly in the same as shown in Fig. \ref{mesh2}(a).

The explicit expression for $\langle W^{s}_{N-q,r}\rangle$ in terms of
$\bar{\Lambda}$'s can be directly deduced from the recurrence
relations for the $\bar{t}$'s.

\subsubsection{Correlations of the type $\langle
W^{s_1}_{N-q_1,r_1}\ldots W^{s_m}_{N-q_m,r_m}\rangle$:} It turns out
that these quantities can be calculated using the results for $\langle
W^{s}_{N-q,r}\rangle$.

Consider the simplest case $\langle
W^{s_1}_{N-q_1,r_1}W^{s_2}_{N-q_2,r_2}\rangle$ with $q_2\geq q_1$.  If
the point $(q_2,r_2)$ does not lie on one of the two lines of defects
originating from $(q_1,r_1)$, it is clear that the effects on the
recurrence relations of the defects originating from $(q_1,r_1)$ and
$(q_2,r_2)$ do not ``interact'' with each other.

The defects do ``interact'' only if $(q_2,r_2)$ lies on the line
defined by $(q_1+k,r_1-k)$ for $0\leq k \leq r_1-1$. In that case, the
defects are of $s_1$-th order of type I on $(q_1+k,r_1-k)$, $0\leq k
\leq r_1-r_2-1$, and of type II on $(q_1+k,r_1-k-1)$, $0\leq k \leq
r_1-r_2-2$; and then of $(s_1+s_2)$-th order of type I on
$(q_1+k,r_1-k)$,$r_1-r_2\leq k \leq r_1-1$. and type II
$(q_1+k,r_1-k-1)$, $r_1-r_2\leq k \leq r_1-2$.

The modifications of the fundamental relations for all the higher
correlations $\langle W^{s_1}_{N-q_1,r_1}\ldots
W^{s_m}_{N-q_m,r_m}\rangle$ can be obtained by using these
observations. Once again if the lines of defects originating from the
points $(q_i,r_i)$ for $i=1,\ldots, m$ do not intersect, the
individual defects do not ``interact''. If they do intersect, the
defects ``interact'' as  described above, implying that all
correlations can be expressed in terms of $\bar{\Lambda}$'s.

\subsection{Comparison with simulation results \label{sec4.5}}

\begin{figure*}
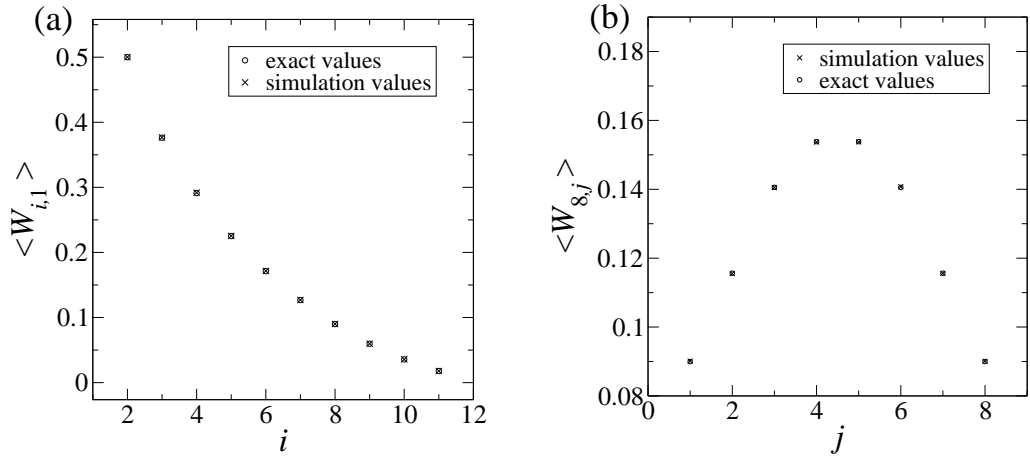

\begin{center}
\begin{minipage}{0.45\linewidth}
\includegraphics[width=0.9\linewidth]{fit_bound}
\end{minipage}
\begin{minipage}{0.45\linewidth}
\includegraphics[width=0.9\linewidth]{fit_bulk}
\end{minipage}
\end{center}
\noindent \caption{Comparison between exact results and simulations:
(a) on the boundary $r=1$ for $N=11$; (b) in the bulk for $q=3$ and
$N=11$.\label{compare}}
\end{figure*}

To evaluate $\langle W_{i,j}\rangle$'s exactly using the exact
relations developed above for any system size, we need to compute the
integers $\Lambda^{(p)}_{i_1\ldots i_p}$ and correspondingly the
$\bar\Lambda^{(p)}_{i_1\ldots i_p}$. These integers are defined
recurrently by the relations (\ref{lambda_form}). Although the relations
(\ref{lambda_form}) are simple sums, the number of terms necessary to
evaluate $\Lambda^{(p)}_{i_1\ldots i_p}$ increases as roughly as $(p!)!$
for large $p$, so that for practical purposes, it is difficult to go
beyond $N=11$. The comparison between the exact evaluation of the $\langle
W_{i,j}\rangle$'s with the corresponding Monte-Carlo simulation results
for $N=11$ is shown in Fig. \ref{compare}.

\subsection{An equivalent combinatorial problem \label{sec4.6}}

We have not been able to find an explicit expression or an asymptotic
formula for $\Lambda^{(p)}_{i_1,\ldots,i_p}$ for large $p$.  However,
after some relabeling, a combinatorial interpretation can be obtained
for  $\Lambda^{(p)}_{i_1,\ldots,i_p}$ and related expressions entering
the formulas for the moments of $W_{i,j}$.

This interpretation goes as follows: we consider maps $h$ which
associate a positive integer $h_{(k,l)}$ to the $(k,l)$-th site of the
portion of the square lattice defined by $1\leq k \leq p$ and $1\leq
l\leq k$ [i.e., the triangle in Fig. \ref{mesh1}(a)].  The integers
$h_{(k,l)}$ are constrained by the following inequalities:
\begin{eqnarray}
h_{(1,1)}=1,\nonumber \\ 1\leq h_{(l,1)}\leq 2, \quad \forall
l=2\ldots p \label{h_def}\label{h_cond1}\nonumber \\ 1\leq
h_{(l,k)}\leq k+h_{(l,k-1)}, \quad \forall l=2\ldots p, k=1\ldots l.
\label{e89}
\end{eqnarray}
The inequalities (\ref{h_def}) are in fact just a re-expression of Eq.
(\ref{ind_usual}). Furthermore, Eq. (\ref{lambda_form}) can then be
rewritten as
\begin{eqnarray}
\hspace{-1cm}\Lambda^{(2)}_{h_{(11)}h_{(12)}} & = &
\delta_{h_{(12)},1} \nonumber\\
\hspace{-1cm}\Lambda^{(p)}_{h_{(p,1)}\ldots h_{(p,p)}} & = &
\delta_{h_{(p,p)},1 }\sum_{h_{(p-1,1)},\ldots,
h_{(p-1,p-1)}}\!\!\!\!\!\!\!\Lambda^{(p-1)}_{h_{(p-1,1)},\ldots,
h_{(p-1)}}\prod_{l=1}^{p-1}\Theta[h_{(p,l+1)}-h_{(p-1,l)}].
\label{lambda_form_h}
\end{eqnarray}

Having iterated the recurrence relation, an equivalent ``explicit''
expression for $\Lambda$'s can be obtained as
\begin{eqnarray}
\Lambda^{(p)}_{h_{(p,1)}\ldots h_{(p,p)}}=\sum_{\{h\}}
\prod_{k=1}^{p}\prod_{l=1}^{k}\Theta[h_{(k,l+1)}-h_{(k-1,l)}]\,,
\label{e91}
\end{eqnarray}
where the sum runs over all maps $h$ satisfying Eq. (\ref{h_def}) with
fixed values of $h_{(p,k)}$ for $k=1,\ldots,p$. The products on the
right hand side are obviously non-zero only for maps $h$ satisfying
\begin{eqnarray}
h_{(k,l)}\leq h_{(k+1,l+1)} \quad \forall k=1\ldots p-1, l=1,\ldots,k.
\label{h_cond2}
\end{eqnarray} 

We now introduce a new symbol $Z_{N-2}$ for $\sum_{j_1,\ldots,j_{N-2}}
\Lambda^{(N-2)}_{j_1,\ldots,j_{N-2}}$, which enters the expression of
the normalization constant $\cal N$. It is then easily seen that
$Z_{N-2}$ is simply the total number of maps $h$ satisfying
inequalities (\ref{h_cond1}) and (\ref{h_cond2}) for $p=N-2$.  In a
similar manner, the quantity
$\sum_{j_1,\ldots,j_{N-2}}\bar{\Lambda}^{(N-2)}_{j_1,\ldots,j_{N-2}}(q,r)$
from (\ref{mean_fin}) can be re-expressed in terms of maps
$\bar{h}(q,r)$ as
$\sum_{\{\bar{h}(q,r)\}}[q+1+\bar{h}_{(N-r,q-1)}-\bar{h}_{(N-r-1,q-1)}]
$ where the maps $\bar{h}(q,r)$ satisfy the inequalities
(\ref{h_cond1}) and (\ref{h_cond2})  except on the line
$(l,k)=(N-r+j,q+j)$ for $0 \leq j \leq r-2$, where they satisfy
\begin{eqnarray}
 1\leq \bar{h}_{(l,k)}\leq k+1+\bar{h}_{(l,k-1)}.
\label{h_cond1*}
\end{eqnarray}

Thus, we finally get
\begin{eqnarray}
\hspace{-1.5cm}\langle
W_{N-q,r}\rangle=\frac{1}{\frac{N(N-1)}{2}+1}\,\frac{\bar{Z}_{N-2}(q,r)}{Z_{N-2}}\,\langle
q+1+\bar{h}_{(N-r,q-1)}-\bar{h}_{(N-r-1,q-1)} \rangle_{\bar{h}(q,r)},
\end{eqnarray}
where $\bar{Z}_{N-2}(q,r)$ is the total number of maps $\bar{h}(q,r)$
for $p=N-2$, and the angular brackets on the r.h.s. denote an average
over all maps. Similar expressions can also be obtained for all higher
moments and correlations.

Interestingly, the original symmetry of the model is not apparent at
all in this formulation. Although the final results, once calculated
[cf. Fig.\ref{compare}], display of course the symmetry, it seems
difficult to show that the underlying discrete problem is indeed
symmetric.

\section{Discussion and Conclusions \label{sec5}}
In this paper, we have studied the response of a hexagonal packing of
rigid, frictionless, massless, spherical grains to a single external
force at the top of it, by supposing all mechanically stable force
configurations equally likely. We have shown that this problem is
equivalent to a correlated $q$-model. Interestingly, while the
conventional $q$-model produces a single-peaked, diffusive response,
our model leads to two sharp peaks on the boundary, i.e., on the two
lattice directions emanating from the point of application of the
external force. For systems of finite size, the magnitude of these
peaks decreases towards the bottom of the packing, while progressively
a broader, central maximum appears between the peaks. The response
function displays a remarkable scaling behaviour with system size $N$:
while the response in the bulk of the packing scales as $\frac{1}{N}$,
on the boundary it is independent of $N$, so that in the thermodynamic
limit only the peaks on the lattice directions persist. We have
obtained exact expressions of the $\langle W_{i,j} \rangle$ values,
i.e., of the response, and of higher correlations for any system size in
terms of integers corresponding to an underlying discrete
structure, but we have not been able to derive an expression for the
scaling limit. The resemblance of the discrete structure with
plane-partition problems \cite{asm} might lead to further progress in
that direction.

Qualitatively, the response obtained via the uniform probability hypothesis is thus in agreement with experiments. It
should however be noted that in experiments, the width of the peaks
increases linearly with depth, while in our model, the peaks are
single-grain diameter wide at any depth. Such peak widening, in our
opinion, is a consequence of inter-grain friction. Indeed, a recent
study \cite{clement} has shown that friction in rectangular packings
produces a widening of the peaks around the two lattice directions
emanating from the point of application of the single external force,
with the peak width proportional to the square root of depth. However,
in Ref. \cite{clement}, the various force configurations have not been
sampled uniformly as in our model, but rather the sampling was carried
out in a fashion similar to the usual $q$-model, with independent
random values at each grain. We are now attempting to study the effect
of friction within the uniform ensemble in both rectangular and
hexagonal geometries; but our preliminary attempts reveal that such a
study is technically difficult as the mapping to an analogous
$q$-formulation breaks down.

A natural question that arises in view of our work is whether the
uniform probability hypothesis leads to one of three categories that
the existing continuum-type models for the transmissions of stresses
have been classified into (namely, elliptic, hyperbolic and parabolic
according to the nature of the underlying coarse-grained PDE's)
\cite{Brev}. Given the critical importance of the underlying network
of contacts, it seems difficult to give a general answer to this
question. For the hexagonal geometry that we studied, it is possible
to write a coarse-grained equation for vertical stresses $\omega$
using the $q$-formulation; indeed, since the correlation length
between the $q$'s is rather short, the continuous limit is the same is
in the usual $q$-model \cite{qresponse}
\begin{eqnarray}
\partial_z \omega + \partial_x (v\omega)=D_0\partial_{xx}\omega\,,
\end{eqnarray}
where $v(x,z)$ is the noise resulting from the continuous limit of
$q_{i,j}=(1+v_{i,j})/2$. However, in contrast with the usual
$q$-model, the mean $\langle v(x,z) \rangle$ is position-dependent as
found in Sec. \ref{sec3.3}. Such a formulation  thus does  not seem
very useful, as the properties of the noise and the self-similarity
have to be somehow first obtained from the equal-probability ensemble
hypothesis. Nevertheless, the study of a tensorial formulation of this
model is an interesting future direction.

Finally, apart from the inclusion of friction in the present model,
the next important future direction is the study of the response of
disordered granular packings. In that case, averages would have to be
taken first for a fixed contact network, and then over different
contact networks corresponding to different arrangement geometries of
grains. The shape of the resulting response function would
provide another crucial test for the applications of the equal
probability hypothesis to the study of forces in granular packings.

\section*{Acknowledgments}
It is a pleasure to thank J.-P. Bouchaud, D. Dhar, J. M. J. van
Leeuwen, B. Nienhuis, J. H. Snoeijer and D. Wolf for useful
discussions. Financial support was provided by the Dutch research
organization FOM (Fundamenteel Onderzoek der Materie).

\end{document}